\documentclass[pre,aps,onecolumn,superscriptaddress]{revtex4}
\usepackage{amsmath,amssymb,graphicx}
\usepackage{epsfig}
\usepackage{graphicx}
\usepackage{textcomp}

\begin{document}
\title{Void formation in diffusive lattice gases}
\author{P. L. Krapivsky}
\affiliation{Physics Department, Boston University, Boston MA 02215, USA}
\affiliation{IPhT, CEA Saclay and URA 2306, CNRS, 91191 Gif-sur-Yvette cedex, France}
\author{Baruch Meerson}
\affiliation{Racah Institute of Physics, Hebrew University of Jerusalem, Jerusalem 91904, Israel}
\author{Pavel V. Sasorov}
\affiliation{Keldysh Institute of Applied Mathematics, Moscow 125047, Russia}

\begin{abstract}
\noindent
What is the probability that a macroscopic void will spontaneously arise, at a specified time $T$, in an initially homogeneous gas? We address this question for diffusive lattice gases, and also determine the most probable density history leading to the void formation. We employ the macroscopic fluctuation theory by Bertini \textit{et al}. and consider both annealed and quenched averaging procedures (the initial condition is allowed to fluctuate in the annealed setting). We show that in the annealed case the void formation probability is given by the equilibrium Boltzmann-Gibbs formula, so the probability is independent of $T$ (and also of the void shape, as only the volume matters). In the quenched case, which is intrinsically non-equilibrium, we evaluate the void formation probability analytically for non-interacting random walkers and probe it numerically for the simple symmetric exclusion process. For voids that are small compared with the diffusion length $\sqrt{T}$, the equilibrium result for the void formation probability  is recovered. We also re-derive our main results for non-interacting random walkers from an exact microscopic analysis.
\end{abstract}

\maketitle
\noindent\large \textbf{Keywords}: \normalsize non-equilibrium processes, large deviations in non-equilibrium systems, stochastic particle dynamics (theory)
\tableofcontents

\section{Introduction}

In thermal equilibrium, fluctuations of macroscopic quantities are completely described in terms of free energy of the system \cite{LLStat}. Non-equilibrium fluctuations are much harder to study, especially when they are large. Anomalously large fluctuations, also called large deviations, of macroscopic observables in \emph{non-equilibrium steady states}  have recently attracted a lot of attention. Several lattice gas models have been investigated and it has been found that the distribution of fluctuations in non-equilibrium steady states, as described by the large deviation functional \cite{Touchette}, can exhibit qualitatively new features, such as non-locality and phase transitions (see reviews \cite{D07,Jona} and references therein).

In this work we examine large deviations which are constrained to occur at a specified moment of time. We are aware of two one-dimensional settings where questions of this type have been already addressed. One of them deals with large fluctuations of integrated current for a step-like initial density profile \cite{DG2009a,DG2009b,varadhan,KM_var}. The other deals with large deviations of dynamical activity, defined as the number of particle moves that have taken place over a given time window in finite systems with exclusion interaction \cite{van}. In general, problems of these type demand an explicit account of the system dynamics.

Here we consider a different type of dynamic problem which allows us to study large deviations, for a whole class of lattice gases,  in arbitrary spatial dimension. Starting with an infinite $d$-dimensional lattice filled with particles at constant density $n$, we study the probability of formation of a void, viz. a large empty region of a given shape with characteristic linear size $L$, at a specified time $T$. A remarkably similar characterization of large deviations has been considered in the context of quantum many-body systems, see \textit{e.g.} \cite{Abanov1,Abanov2} and references therein, under the name of ``emptiness formation".

We will study the probability of void formation in two different settings --- starting from a fluctuating equilibrium state (the so-called annealed setting) or from a deterministic state (the quenched setting). In the annealed setting the gas remains in equilibrium, for any $T$, in the process of void formation, as we explicitly show in this work. The quenched setting is intrinsically non-equilibrium, since it describes the void formation process as the gas is evolving on its way to equilibrium.

Our analysis is based on a coarse-grained approach which directly probes the long-time limit and is applicable to a whole family of diffusive lattice gases of interacting particles. This approach is rooted in the \emph{macroscopic fluctuation theory} (MFT) of Bertini, De Sole, Gabrielli, Jona-Lasinio, and Landim \cite{Bertini}, see also Refs. \cite{Tailleur,DG2009b,van}. The MFT is a generalization of the low-noise Freidlin-Wentzell theory \cite{FW84} which in turn is an analog of the WKB approximation of quantum mechanics. Similar approaches have been also developed for lattice gases where, in addition to diffusive transport, there are on-site reactions among particles \cite{EK,MS2011,MSfronts}. In its standard form, the MFT \cite{Bertini} holds for lattice gases where the transport is unbiased and diffusion-like. In addition to non-interacting random walks (RWs), well-known examples of such lattice gases are the simple symmetric exclusion process (SSEP) \cite{Spohn,KL99,L99,SZ95,D98,S00,BE07,KRB10}, the Kipnis-Marchioro-Presutti (KMP) model \cite{KMP,BGL,van_2}, and the symmetric zero range process (ZRP)  \cite{Spitzer,Spohn,Evans}. In the SSEP a particle can hop to a neighboring site if that site is empty; if it is occupied by another particle, the move is disallowed. The KMP model describes a lattice of mechanically uncoupled harmonic oscillators which randomly redistribute energy among neighbors. The ZRP describes \emph{interacting} random walks: A particle at site $\mathbf{i}$ can hop to a neighboring site with rate $\alpha$ that depends on the number of particles $n_\mathbf{i}$ on the departure site $\mathbf{i}$. For these and other similar lattice gases, a hydrodynamic, or mean-field, description is provided by a diffusion equation
\begin{equation}
\label{rho:eq}
\partial_t \rho=\nabla \cdot [D(\rho) \nabla \rho]
\end{equation}
for the average density $\rho(\mathbf{r},t)$  \cite{monotone}.  The diffusion coefficient $D(\rho)$ is constant in the simplest models (\textit{e.g.}, for the RWs, the SSEP, and KMP), but generally it depends on the density; hence the diffusion equation (\ref{rho:eq}) is generally non-linear.

At the level of MFT, fluctuating diffusive gases are fully characterized by $D(\rho)$ and an additional  function $\sigma(\rho)$ which describes equilibrium fluctuations \cite{Bertini,Spohn}. Table I lists the functions $D(r)$ and $\sigma(r)$ for the four aforementioned models: the RWs,  the SSEP, the KMP and the ZRP (in the latter case we assume that $\alpha(r)$ is a monotonically increasing function \cite{ZRPproof}). It also gives, for these models, the equilibrium free energy density $F(r)$. The free energy density is related to $D(r)$ and $\sigma(r)$:
\begin{equation}
\label{FDT}
F^{\prime\prime}(r) = \frac{2D(r)}{\sigma(r)}\,.
\end{equation}
(Hereinafter, the prime denotes the derivative.) Equation~(\ref{FDT}) follows from the fluctuation-dissipation theorem  \cite{Spohn,D07}, and it also naturally emerges from the MFT formalism \cite{KM_var}.

\begin{table}[t]
\label{Table_models}
\begin{tabular}{|c|c|c|c|}
\hline
 ~Model ~ &  ~$D(r)$ ~ & $\sigma(r)$ & ~$F(r)$  \\
\hline
RWs      &  1    &      $2r$             &  $r\ln r - r$\\
\hline
SSEP     &  1     &  ~$2r(1-r)$ ~    &   ~$r\ln r  +(1-r)\ln(1-r)$ ~\\
\hline
KMP     &  1     &   ~$4r^2$ ~     &   ~$-(1/2)\,\ln r $ ~\\
\hline
ZRP     &  $\alpha^{\prime}(r)$     &   ~$2 \alpha(r)$ ~   &   ~$\int^r du \, \ln \alpha(u)$ ~\\
\hline
\end{tabular}
\caption{Functions $D(r), \sigma(r)$ and  $F(r)$ for non-interacting random walkers (RWs), the SSEP, the KMP
and the ZRP models.}
\end{table}

In this paper we derive and analyze the MFT equations and boundary conditions describing formation of a void in a whole class of diffusive gases in arbitrary spatial dimension. We show that, for these systems, the probability ${\mathcal P}$ of void formation has a universal scaling form: $\ln {\mathcal P} \simeq -\,T^{d/2} \mathcal{S}(L/\sqrt{T},n)$. In the annealed setting, the large deviation function $\mathcal{S}$ is such that the resulting ${\mathcal P}$ is independent of $T$ and described by the classical Boltzmann-Gibbs equilibrium formula.  Correspondingly, the optimal (most probable) time history of the gas density field in the process of void formation coincides in this case with a time-reversed history of the mean-field relaxation process, that is with a time-reversed solution of the diffusion equation.

Finding $\mathcal{S}$ analytically in the quenched setting is a hard problem, since the optimal time history of the gas density in the process of void formation is different from the time-reversed solution of the diffusion equation. We have only been able to solve this problem analytically for non-interacting RWs. In general, $\mathcal{S}$ depends on the void shape.  For short times (equivalently, large voids), the large deviation function is a genuinely non-equilibrium quantity that admits an integral representation with an interesting geometric flavor: $\mathcal{S}\simeq n \int_\mathcal{V} d{\mathbf X}\,\, [\mathcal{D}({\mathbf X})]^2$. Here $\mathcal{V}$ is obtained by rescaling all the coordinates of the void by the characteristic diffusion length $\sqrt{4T}$, and $\mathcal{D}({\mathbf X})$ is the distance between the point $\mathbf{X}$ inside the rescaled void $\mathcal{V}$ and its boundary $\partial\mathcal{V}$. In particular, this implies that the spherical void is the \emph{least probable} among voids of the same volume.  For long times (equivalently, small voids), the large deviation function $\mathcal{S}$ becomes  shape-independent, and the equilibrium result  for ${\mathcal P}$ is recovered. The equilibrium result is also obtained when $\max \limits_{\mathbf{X}} \mathcal{D}({\bf X})\ll 1$.

Even in the relatively simple case of non-interacting RWs, the MFT formalism turns out to
be quite instructive. Not only it accurately predicts the logarithm of the void formation probability, but it also gives the optimal time history of the gas density field. This history strongly depends on the ratio $L/\sqrt{T}$.  For interacting lattice gases, the void formation probability, and the optimal density history, can be found by solving the MFT equations numerically, as we demonstrate for the SSEP.

The remainder of the paper is structured as follows. In section \ref{MFT} we consider a diffusive gas with arbitrary $D(q)$ and $\sigma(q)$ and present the MFT equations and boundary conditions for the void formation problem; details of the derivations are given in Appendix A. In section \ref{annealed} we solve the void formation problem in the annealed setting. Sections \ref{RW} and \ref{num} deal with the quenched setting: Analytical results for the RWs are established in section \ref{RW}, while numerical results for the SSEP are given in section \ref{num}.  Our main findings are briefly discussed in section \ref{summary}. In Appendix B we outline an exact \emph{microscopic} theory of the void formation for the RWs, both in the quenched and annealed settings. This microscopic theory yields the void formation probability and the expected system configuration at $t=T$ which coincide, in the long time limit, with the corresponding results obtained from the MFT formalism.

\section{Macroscopic fluctuation theory  of void formation}
\label{MFT}

\subsection{Governing equations and boundary conditions}
\label{equation}

Our analysis employs the MFT  \cite{Bertini,Tailleur,DG2009b}:  a coarse-grained formalism which directly probes the long-time limit and is valid for a whole family of diffusive lattice gases. The MFT can be formulated as a classical Hamiltonian field theory, where the number density $q(\mathbf{x},t)$ plays the role of ``coordinate", and the conjugate field $p(\mathbf{x},t)$ (which can be viewed as the magnitude of fluctuations)  is the ``momentum". In Appendix A we present a derivation of the Hamilton equations and boundary conditions for the void formation problem in the annealed and quenched settings.  The derivation starts from \emph{fluctuating hydrodynamics}: a Langevin-type partial differential equation for $q(\mathbf{x},t)$ whose deterministic part coincides with Eq.~(\ref{rho:eq}), whereas the (multiplicative) noise term includes $\sigma(q)$ \cite{Spohn}. Being interested in the long-time behavior of the probability distribution, we arrive at a variational problem (see Appendix A) which leads to
two coupled partial differential equations for $q(\mathbf{x},t)$ and $p(\mathbf{x},t)$:
\begin{eqnarray}
  \partial_t q &=& \nabla \cdot \left[D(q) \nabla q-\sigma(q) \nabla p\right], \label{d1} \\
  \partial_t p &=& - D(q) \nabla^2 p-\frac{1}{2} \,\sigma^{\prime}(q) (\nabla p)^2. \label{d2}
\end{eqnarray}
Equations~\eqref{d1} and \eqref{d2} are Hamiltonian, since they can be written as
\begin{equation}
\partial_t q = \delta H/\delta p\,, \quad
\partial_t p = -\delta H/\delta q\,.
\end{equation}
Here
\begin{equation}
\label{Hamiltonian}
H[q(\mathbf{x},t),p(\mathbf{x},t)]= \int d\mathbf{x}\,\mathcal{H}
\end{equation}
is the Hamiltonian,
\begin{equation}
\label{Ham}
\mathcal{H}(q,p) = -D(q) \nabla q\cdot \nabla p
+\frac{1}{2}\sigma(q)\!\left(\nabla p\right)^2.
\end{equation}
The spatial integration in Eq.~\eqref{Hamiltonian}, as well as in a number of equations below, is over the entire space.  For a given lattice gas model, specified by $D(q)$ and $\sigma(q)$, the same Eqs.~\eqref{d1} and \eqref{d2} arise when one studies large deviations of different quantities in different settings. Boundary conditions in space and time, that complement Eqs.~\eqref{d1} and \eqref{d2}, vary from problem to problem.  Here we study the formation of a void in an infinite system. Hence we demand that, at a specified time $t=T$, a (simply connected) void of a given shape is observed:
\begin{equation}
\label{void}
q(\mathbf{x},T)=0\;\;\;\text{inside the void}.
\end{equation}
We emphasize that we do not specify the density profile $q(\mathbf{x},T)$ outside the void --- it will emerge from the solution of the problem as the density profile that \emph{maximizes} the probability to observe the void (\ref{void}).

The void formation probability not only depends on the dynamics of the underlying microscopic model during the time interval $0<t<T$, but also on the initial condition. At the macroscopic level we want the initial density to be uniform. One way to achieve it is to start,  in the microscopic formulation, with a \emph{deterministic} constant density. The void formation probability is then obtained by averaging only over stochastic realizations of the dynamics over the time interval $0<t<T$. Alternatively, we can allow equilibrium fluctuations of the initial condition and average over both these fluctuations and stochastic realizations of the dynamics. These two types of averaging are called quenched and annealed, respectively \cite{DG2009b}. This terminology suggests an analogy with the quenched and annealed averaging in disordered systems, although there is no disorder in the present situation. Similarly to disordered systems, the analysis tends to be simpler in the annealed case, as we will see shortly.  

We now summarize the rest of boundary conditions for the MFT equations \eqref{d1} and \eqref{d2}.
As shown in Appendix A, the maximization of the void formation probability yields the following boundary condition at $t=T$:
\begin{equation}
\label{pT}
p(\mathbf{x},T)=0\;\;\;\text{outside the void},
\end{equation}
for both annealed and quenched settings.  Essentially, Eq.~(\ref{pT}) states that fluctuations outside the void at $t=T$ must vanish.
The boundary condition at $t=0$ does depend on the setting. For the quenched setting the initial condition is
\begin{equation}
\label{flat}
    q(\mathbf{x},0)=n\;\;\;\text{everywhere}.
\end{equation}
The annealed setting assumes equilibrium fluctuations in the initial condition, i.e., the density profile  at $t=0$ is chosen from the equilibrium probability distribution corresponding to density $n$. For the annealed setting the initial condition for Eqs.~\eqref{d1} and \eqref{d2},
\begin{equation}\label{annealed0}
p(\mathbf{x},0)=\mathcal{F}^{\prime}[q(\mathbf{x},0)],
\end{equation}
establishes a relation between the most probable initial density profile and the corresponding profile of $p(\mathbf{x},0)$, see Appendix A. The function $\mathcal{F}(r)$ is simply related to the free energy density $F(r)$, differing from it by a linear function. Namely, $\mathcal{F}(r)$ obeys Eq.~(\ref{FDT}) and the additional relation $\mathcal{\mathcal{F}}^{\prime}(n)=0$. For concreteness, we also set $\mathcal{\mathcal{F}}(n)=0$, so that
\begin{equation}\label{modifiedF}
    \mathcal{F}(r) = \int_n^r d\xi \int_n^{\xi} d\zeta \,\frac{2 D(\zeta)}{\sigma(\zeta)}.
\end{equation}
Finally, the boundary conditions at $\mathbf{x}\to \infty$ are
\begin{equation}\label{infinity}
q(\mathbf{x}\to \infty,t)=n,\;\;\;p(\mathbf{x}\to \infty,t)=0.
\end{equation}
Although Eqs.~\eqref{d1} and \eqref{d2} and the boundary conditions represent a complete set for the quenched and annealed settings, there is an important corollary [valid under certain conditions on the
functions $D(r)$ and $\sigma(r)$] that can be very useful for solving the problem. This corollary is
\begin{equation}
\label{pT2}
p(\mathbf{x},T)=-\infty\;\;\;\text{inside the void},
\end{equation}
in both quenched and annealed settings, see Appendix A.

One additional comment is in order about the fields $q$ and $p$ at $t=0$ in the quenched setting. Although the initial density profile $q(\mathbf{x},0)=n$ is flat here, the ({\it a priori} unknown)
initial momentum  $p(\mathbf{x},0)$
is nonzero. Therefore, the initial state, as described by the MFT, is non-deterministic. How to reconcile this fact with our definition of the quenched setting as the one starting from a deterministic density profile? The solution comes from the realization that, for the MFT formalism to become valid, one should wait for a time which is much longer than the characteristic microscopic time of the system: the time of a single particle move. During this time (which is assumed to be very short compared to the macroscopic time scales that the MFT formalism can only deal with) the deterministic density profile rapidly evolves into a fluctuating profile with constant average density $n$. In contrast to the annealed setting, however, these fluctuations are not in equilibrium: they are determined by the \textit{a priori} unknown field $p(\mathbf{x},0)$ that depends on $\ell$ and is different from the equilibrium $p$-field described by Eq.~(\ref{annealed0}).

Solutions of Eqs.~(\ref{d1}) and (\ref{d2}) with vanishing momentum, $q(\mathbf{x},t)=\rho(\mathbf{x},t)$ and $p(\mathbf{x},t)=0$, are deterministic; they are called relaxation solutions \cite{gradp}. For the relaxation solutions Eq.~\eqref{d2} is satisfied identically, whereas Eq.~\eqref{d1} reduces to the diffusion equation \eqref{rho:eq}. Solutions with $p(\mathbf{x},t)\neq 0$ are called activation solutions, for these solutions $q(\mathbf{x},t)\neq \rho(\mathbf{x},t)$. The void formation obviously
demands an activation solution. Once the activation solution $q(\mathbf{x},t)$ and $p(\mathbf{x},t)$, obeying the boundary conditions,  is found, we can evaluate  the action of the Hamiltonian system (\ref{d1}) and (\ref{d2}):
\begin{equation}
\label{action}
    S =
    \int d\mathbf{x} \int_0^T dt\, \left(p \partial_t q -\mathcal{H}\right)
    = \frac{1}{2}\int d\mathbf{x} \int_0^T dt\, \sigma(q) \,(\nabla p)^2.
\end{equation}
For the quenched setting, this action yields the void formation probability:  ${\ln \mathcal P} \simeq - S$.
For the annealed setting, one also has to account for the ``cost" $S_0$ of creating the optimal initial density profile $q(\mathbf{x},0)$ in the equilibrium gas of average density $n$. This cost is given by the Boltzmann-Gibbs equilibrium formula, so
\begin{equation}
\label{S0}
    S_0=
    \int d\mathbf{x}\,\mathcal{F}[q(\mathbf{x},0)]
\end{equation}
(recall that we defined $\mathcal{F}(r)$ so that $\mathcal{F}(n)=0$.) Therefore, for the annealed setting we have ${\ln \mathcal P} \simeq - (S_0+S)$.

\subsection{Dynamic scaling of the void formation probability}
\label{scaling}

For a given diffusive lattice gas, the action $S$ depends on the characteristic size of the void $L$, the formation time $T$, and the gas density $n$. Let us rescale time by $T$, $t/T \to t$, and the distances by the diffusion length, $\mathbf{x}/\sqrt{T} \to \mathbf{x}$. Equations (\ref{d1}) and (\ref{d2}) remain invariant under this transformation. The boundary conditions (\ref{void}) and (\ref{pT}) [and the corollary (\ref{pT2})] remain the same except that $T$ is replaced by 1, and the physical void is replaced by the void of rescaled size $L/\sqrt{T}$. The rest of the boundary conditions do not change. Equation~(\ref{action}) becomes
\begin{equation}
\label{actionrescaled}
   S(L,T,n)=\frac{1}{2}\,T^{d/2} \int d\mathbf{x} \int_0^1 dt\, \sigma(q) \,(\nabla p)^2,
\end{equation}
and Eq.~(\ref{S0}) also acquires factor $T^{d/2}$. Now, parameters $L$ and $T$ only enter the rescaled problem via the combination $L/\sqrt{T}$. Therefore $p(x,t)$ and $q(x,t)$ may depend on $L$ and $T$ only via this combination. This implies a scaling form
\begin{equation}
\label{scalingform}
    S(L,T,n) = T^{d/2} \mathcal{S}(L/\sqrt{T},n)
\end{equation}
in both quenched and annealed settings, and for a whole class of symmetric diffusive lattice gases.  The problem, therefore, is reduced to finding the large deviation function $\mathcal{S}(L/\sqrt{T},n)$ which coincides with the mechanical action of the rescaled
problem. In the next section we show that in the annealed setting the large deviation function drastically simplifies, viz. the sum $S_0+S$ is given by the Boltzmann-Gibbs equilibrium formula, so the void formation probability is independent of $T$ and of the void shape.

\section{Void formation in annealed setting}
\label{annealed}

\subsection{General}
\label{genannealed}

The annealed setting turns out to be simple. The reason is that the initial condition (\ref{annealed0}) belongs to the invariant \emph{equilibrium manifold} of Eqs.~(\ref{d1}) and (\ref{d2}),
described by the local relation
\begin{equation}\label{local}
p(\mathbf{x},t)=\mathcal{F}^{\prime}[q(\mathbf{x},t)]
\end{equation}
between $q$ and $p$ \cite{KM_var}. Using this relation along with
Eq.~(\ref{FDT}),  one can see by direct calculation that $q(x,t)$ is governed by the \emph{time-reversed} deterministic equation
\begin{equation}
\label{antidiff}
\partial_t q=-\nabla \cdot [D(q) \nabla q],
\end{equation}
as expected for reversible models, like ours, in equilibrium. Using Eqs.~(\ref{void}), (\ref{pT}) and (\ref{local}), we obtain the full density profile at $t=T$:
\begin{equation}
\label{qfullT}
q(\mathbf{x},T) =
\begin{cases}
0,   & \text{inside the void},\\
n,  & \text{outside the void}.
\end{cases}
\end{equation}
With this condition we can solve the ``anti-diffusion" equation (\ref{antidiff})
backward in time.
In this way we obtain the optimal density history of the gas, whereas $q(x,0)$ yields the optimal initial condition.
Now we can calculate $p(x,t)$ from the local relation (\ref{local}) and determine $S$ and $S_0$ from Eqs.~(\ref{action}) and (\ref{S0}). The following shortcut, however, makes these calculations redundant. The creation of optimal initial profile $q(\mathbf{x},0)$  at $t=0$,  followed by the void formation at $t=T$,  can be described as a single \emph{extended} activation trajectory $q(\mathbf{x},t)$ that starts, at $t=-\infty$, from the flat state $q(\mathbf{x},t=-\infty)=n$, acquires the optimal shape $q(\mathbf{x},0)$ at $t=0$ and ends by forming the desired void at $t=T$. This extended trajectory
belongs to the invariant equilibrium manifold. Therefore, the cost of creating a void is determined by the Boltzmann-Gibbs equilibrium formula, and we obtain
\begin{eqnarray}
\label{annealedP}
\ln \mathcal{P}_{\text{annealed}} &\simeq&  -
    \int d\mathbf{x}\,\mathcal{F}[q(\mathbf{x},T)]
   = -\int\limits_{\text{void}} d\mathbf{x}\,\mathcal{F}(0) = -\mathcal{F}(0) V,
\end{eqnarray}
where $V$ is the volume of the void, and we have used our convention $\mathcal{F}(n)=0$, see Eq.~(\ref{modifiedF}).  The
equilibrium result (\ref{annealedP}) is independent of $T$ and of the void shape.

Table II yields $\ln \mathcal{P}$ for the models listed in Table \ref{Table_models}.
Note that $\ln \mathcal{P}$ is proportional to $n$ for the RWs reflecting their non-interacting character.
For the KMP model, $\mathcal{F}(0)=\infty$ which implies zero void formation probability; this is also evident from the definition of the microscopic model \cite{KMP}. For the SSEP, $\mathcal{F}(0)$ diverges as $n\to 1$, again as expected from the microscopic model.

\begin{table}
\label{Table_annealed}
\begin{tabular}{|c|c|c|}
\hline
 ~Model ~ &  ~$\ln \mathcal{P}_{\text{annealed}}$ ~\\
\hline
RWs      &  $-V n$   ~\\
\hline
SSEP     &  $-V \ln\frac{1}{1-n}$  ~\\
\hline
KMP     &  $-V \times \infty = -\infty$  ~\\
\hline
ZRP     &  $-V \left[n \ln n-\int_0^n du \ln \alpha(u)\right]$  ~\\
\hline
\end{tabular}
\caption{$\ln \mathcal{P}$ in the annealed setting for the RWs, the SSEP, the KMP, and the ZRP. The results depend only on the average gas density $n$ and the volume of the void $V$.}
\end{table}

Although the probability $\mathcal{P}$ depends, via $\mathcal{F}(0)$,  on both $D(r)$ and $\sigma(r)$, the density histories only depend on $D(r)$ but not on $\sigma(r)$. In other words, all diffusive lattice gases with the same $D(r)$ and the same average density have identical optimal density histories of void formation in the annealed setting. We now present more details on the void formation in one dimension.

\subsection{Constant diffusion coefficient}
\label{ldannealed}

\begin{figure}
\includegraphics[width=2.5 in,clip=]{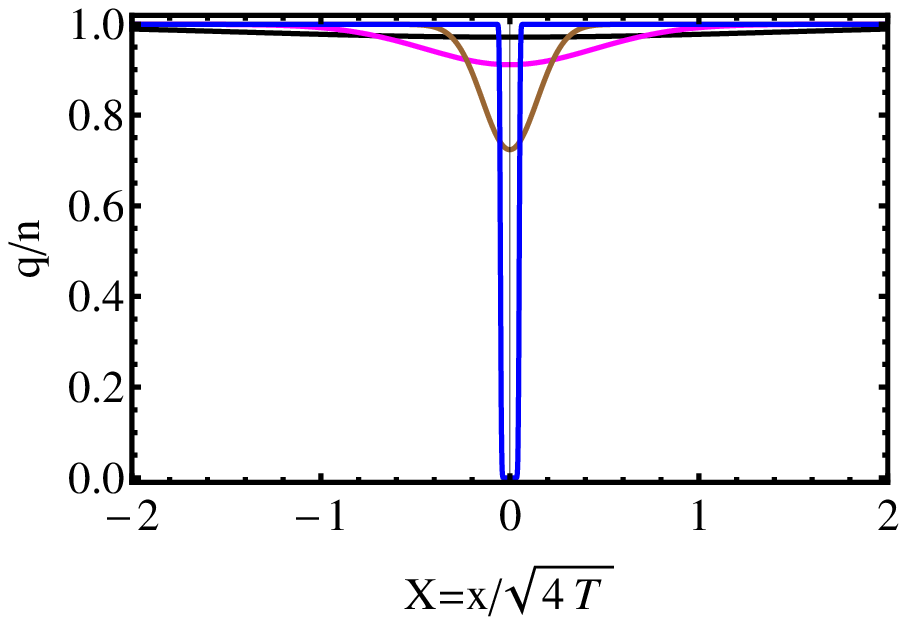}
\includegraphics[width=2.5 in,clip=]{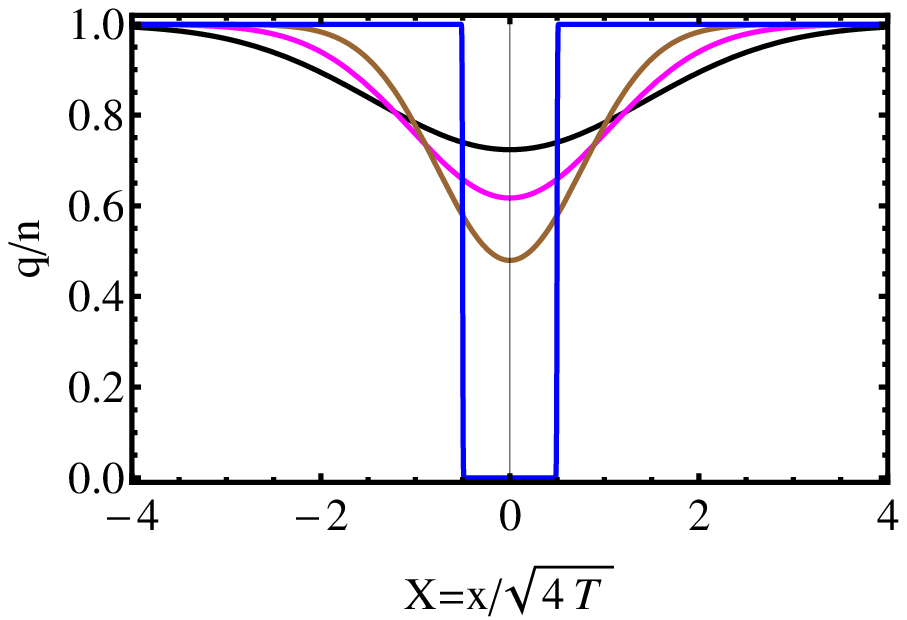}
\includegraphics[width=2.5 in,clip=]{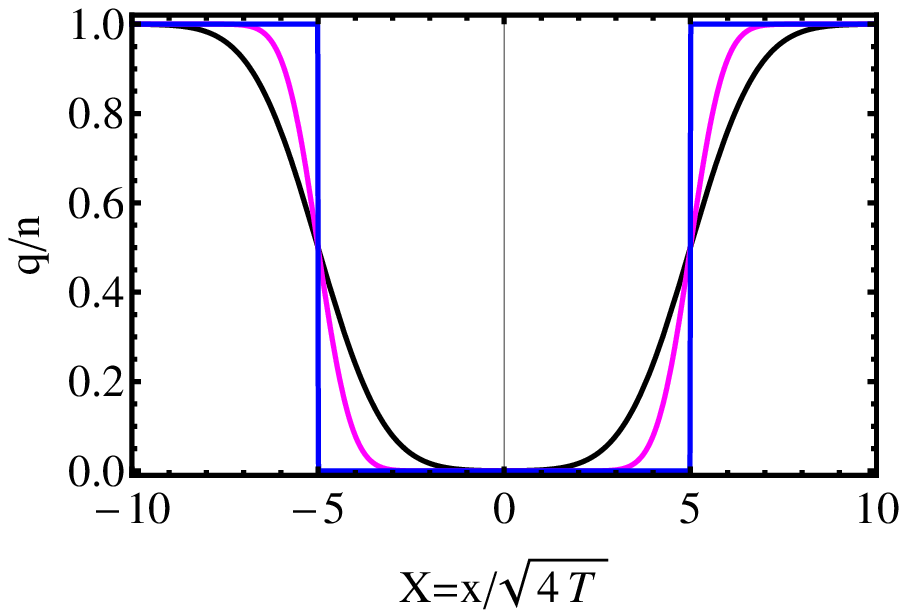}
\caption{(Color online)} Void formation in the annealed case for models with $D=1$. Shown are the optimal density profile histories $q(X,t)/n$ at times $t/T=0, 0.9, 0.99$ and $1$ for  $\ell=L/\sqrt{4T}=0.05$ (left panel), at times $t/T=0, 0.5, 0.75$ and $1$ for $\ell=0.5$ (right panel), and at times  $t/T=0, 0.75$ and $1$ for  $\ell=5$ (bottom panel). The activation solutions are time-reversed relaxation solutions.
\label{anvoid}
\end{figure}

Let $D(q)=1$. Such a density-independent diffusion coefficient characterizes, \textit{e.g.}, the RWs, the SSEP, and the KMP. In one dimension, the void is a segment (we always tacitly assume that the void is a connected set). We can set the void to be the $[-L,L]$ segment. Thus Eq.~(\ref{qfullT}) becomes
\begin{equation}
\label{q(x,T)}
q(x,T)=n \,H(|X|-\ell) \equiv n \,H(|x|-L),
\end{equation}
where $H(z)$ is the Heaviside step function.  The optimal density history of void formation  is described by the solution of the linear anti-diffusion equation:
\begin{equation}
\label{q(x,t)}
    q(x,t)=\frac{n}{2}\,\text{erfc}\left(\frac{\ell-X}{
   \sqrt{1-t/T}}\right)+\frac{n}{2}\,\text{erfc}\left(\frac{\ell+X}{\sqrt{1-t/T}}\right),
\end{equation}
where $X=x/\sqrt{4T}$, $\ell =L/\sqrt{4T}$ and $\text{erfc}(z)=(2/\sqrt{\pi}) \int_z^{\infty} \exp(-\xi^2) d\xi$ is  the complementary error function. The optimal initial density profile, therefore, is
\begin{equation}
\label{q(x,0)}
    q(x,0)=\frac{n}{2}\,\text{erfc}\left(\ell-X\right)+
   \frac{n}{2}\,
   \text{erfc}\left(\ell+X\right).
\end{equation}
Some examples of the density history are shown on Fig.~\ref{anvoid}. The left panel corresponds to small $\ell$. In this situation, the optimal initial profile $q(x,0)$ is almost flat. As a result, most of the actual void formation occurs towards the end of the time interval $0<t<T$. The bottom panel shows the opposite regime (large $\ell$). Here the optimal initial density profile already has a pronounced dip: the equilibrium fluctuations had to do most of the job already at $t<0$.

Equations~(\ref{annealedP}) and (\ref{q(x,0)}) for the RWs also follow from exact results of the  microscopic theory (Appendix B).

\subsection{Nonlinear diffusion}
\label{nldannealed}

For a density-dependent diffusion coefficient, the optimal density histories [described by the nonlinear anti-diffusion equation (\ref{antidiff}) with the boundary condition (\ref{q(x,T)})] are more interesting. One fascinating phenomenon appears in a class of models satisfying $D(0)=0$. In such models,  in the process of void formation, the gas density vanishes in a finite region of space already at an earlier time $\tau<T$. The empty region grows with time until the complete void (\ref{q(x,T)}) is formed at $t=T$. For sufficiently small $L/\sqrt{T}$ one has $\tau>0$, and the initial density $q(\mathbf{x},0)$ is everywhere positive. Otherwise the empty region is already present in the optimal initial density $q(\mathbf{x},0)$.

\begin{figure}
\includegraphics[width=2.5 in,clip=]{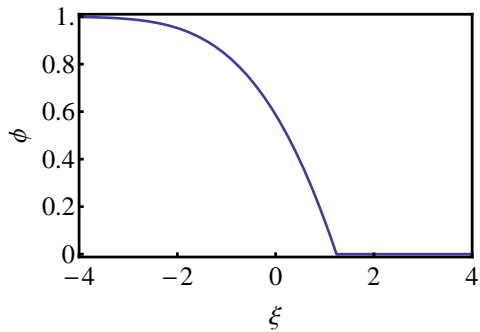}
\caption{(Color online)} The numerically found shape function $\phi_{k=1}(\xi)$ of the similarity solution (\ref{ss}).
\label{compact_ODE}
\end{figure}

We now give more details in the case when the diffusion coefficient has a simple algebraic form: $D(q)=q^k$ with $k>0$. When $L/\sqrt{T}<\xi_k$, the initial density is everywhere positive. If $L/\sqrt{T}>\xi_k$, the initial density $q(x,0)$ vanishes on a finite interval. The factor $\xi_k$ comes from the self-similar solution,
\begin{equation}
\label{ss}
q_s(x,t)=n \,\phi\left(\frac{x}{n^{k/2} t^{1/2}}\right),
\end{equation}
of an auxiliary relaxation problem:
\begin{equation}
\label{k=1}
\partial_t q=\partial_x (q^k\partial_x q)
\end{equation}
on the interval $|x|<\infty$, with the initial condition being a step function: $q(x,0)=n H(-x)$. A distinctive feature of this class of problems (see \textit{e.g.} \cite{Zeld,LL87,B96, pablo}) is the (semi-)compact support: there exists a finite point in space, $x_k(t)=\xi_k\,\sqrt{t}$, so that the solution $q_s(x,t)$ is positive at $x<x_k(t)$ and zero at $x\geq x_k(t)$. The shape function $\phi(\xi)$ solves the ordinary differential equation
\begin{equation}\label{ode}
\left(\phi^k \phi^{\prime}\right)^{\prime}+(1/2) \,\xi \phi^{\prime}=0
\end{equation}
with boundary conditions $\phi(\xi\to -\infty)=1$ and $\phi(\xi\to \infty)=0$. This problem can be easily solved numerically by a shooting method. Figure~\ref{compact_ODE} gives an example of numerical solution for $k=1$, that is $D(q)=q$. Here $\xi_{k=1}\simeq 1.239$.

Now consider Eq.~(\ref{k=1}) when the initial condition includes two step functions, as described by the right hand side of Eq.~(\ref{q(x,T)}). Because of the semi-compact support of the similarity solution (\ref{ss}),  the solution of this problem, at sufficiently short times $t<\tau$,  is a sum of two counter-propagating similarity solutions of the type (\ref{ss}):
\begin{equation}\label{two}
q(x,t)= n \,\phi\left(\frac{L+x}{n^{k/2} t^{1/2}}\right)+ n \,\phi\left(\frac{L-x}{n^{k/2} t^{1/2}}\right),\;\;\;t<\tau.
\end{equation}
The character of solution changes when the edge points $x_k^{-}=-L+\xi_k \sqrt{t}$ and $x_k^{+}=L-\xi_k \sqrt{t}$ of the two similarity solutions meet. This occurs at $t=L^2/\xi_k^2$. As the activation trajectory is a time-reversed relaxation trajectory, one can easily obtain the condition $L/\sqrt{T}<\xi_k$ for the positiveness of $q(x,0)$ everywhere, and $L/\sqrt{T}>\xi_k$ for the presence of an empty region already at $t=0$. Again, the same density histories will be observed for all models with the same $D(q)$, independently of $\sigma(q)$.

\begin{figure}
\includegraphics[width=2.5 in,clip=]{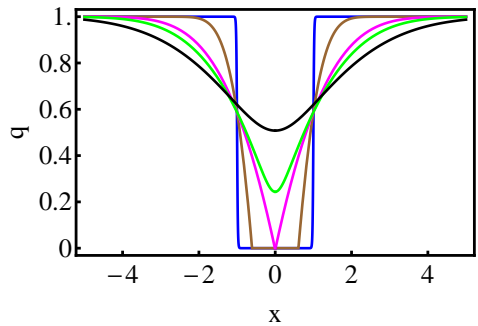}
\caption{(Color online)} Void formation in the annealed case for models with $D(q)=q$. The parameters are $n=1$, $L=1$ and $T=2$. Shown is the optimal density history $q(x,t)$ at times $t=0, 1, 1.35, 1.9$ and $2$. At time $t\simeq 1.35$ an empty interval appears. The activation solutions are time-reversed relaxation solutions.
\label{compact}
\end{figure}

Figure \ref{compact} shows a complete density history for $L=1$ and $T=2$, obtained by solving numerically the anti-diffusion equation (\ref{antidiff}), for $k=1$, with the boundary condition (\ref{q(x,T)}).  As $L/\sqrt{T}<\xi_1$ in this example, the initial density is everywhere positive. An empty interval appears at time $\tau=T-L^2\xi_{k=1}^{-2}\simeq 1.35$ via a corner in the density profile. Finally, $\ln \mathcal{P}_{\text{annealed}}\simeq -2L \times 2n=-2L \times 2\times 1 =-4L$ in this case, see Table 2.  For
comparison, $\ln \mathcal{P}_{\text{annealed}}\simeq -2L$ for the RWs with the same average density $n=1$.

\section{Void formation in quenched setting: non-interacting random walkers}
\label{RW}

\subsection{General}

In the quenched setting one needs to solve Eqs.~(\ref{d1}) and (\ref{d2}) with the boundary conditions
at $t=0$ and $T$ described by Eqs.~(\ref{pT}), (\ref{flat}) and (\ref{pT2}). Now the system does not lie in the invariant equilibrium manifold (\ref{local}),  and the problem of void formation appears intractable for a generic lattice gas. We have succeeded in solving it analytically only for RWs, where Eqs.~(\ref{d1}) and (\ref{d2}) become
\begin{eqnarray}
  \partial_t q &=& \nabla \cdot \left(\nabla q-2q \nabla p\right), \label{rw1} \\
  \partial_t p &=& - \nabla^2 p-(\nabla p)^2. \label{rw2}
\end{eqnarray}
The solution can be obtained via the Hopf-Cole
canonical transformation defined by the relations $Q=q e^{-p}$ and $P=e^p$, see \textit{e.g.} \cite{EK}. The generating function of this
transformation can be chosen as $\int d\mathbf{x} \,\Phi(q,P) =\int  d\mathbf{x} \,  q \ln P$.
The new Hamiltonian is $\int d\mathbf{x} \,\widetilde{\mathcal{H}}$, with density
$\widetilde{\mathcal{H}}=-\nabla Q \cdot \nabla P$. The Hamilton equations become
\begin{eqnarray}
  \partial_tQ &=& \nabla^2 Q, \label{Qt}\\
  \partial_tP &=& -\nabla^2 P. \label{Pt}
\end{eqnarray}
The mechanical action along an activation trajectory can be written as
\begin{eqnarray}
  S &=& \int d\mathbf{x} \int_0^T dt\, \left(p\partial_t q-\mathcal{H}\right) \nonumber \\
    &=& \int d\mathbf{x} \int_0^T dt\, \left(P\partial_t Q-\tilde{\mathcal{H}}\right)+\int d\mathbf{x} \,\Phi(q,P)\,\biggl|_0^T \nonumber \\
  &=& \int_0^T dt \int d\mathbf{x}\,(P \nabla^2 Q+\nabla Q \cdot \nabla P)+ \int d\mathbf{x} \, \Phi(q,P)\,\biggl|_0^T.
\end{eqnarray}
The integral over $\mathbf{x}$ in the first term vanishes by virtue of Green's first identity and the boundary conditions at $\mathbf{x}\to \infty$. As a result, the action
\begin{equation}\label{SRW}
    S=\int d\mathbf{x} \, \Phi(q,P)\,\biggl|_0^T = \int d\mathbf{x} \,\left[q(\mathbf{x},T)\, \ln P(\mathbf{x},T) - q(\mathbf{x},0)\, \ln P(\mathbf{x},0)\right]
\end{equation}
is fully determined by the initial ($t=0$) and final ($t=T$) states. In view of Eqs.~(\ref{pT}) and (\ref{pT2}) we have
\begin{equation}
\label{Td}
P(\mathbf{x},T) =
\begin{cases}
0  & \text{inside the void},\\
1  & \text{outside the void}.
\end{cases}
\end{equation}
Using Eqs.~(\ref{void}) and (\ref{Td}), we can reduce Eq.~(\ref{SRW}) to
\begin{equation}\label{SRW1}
    S= - n \int d\mathbf{x}\, \ln P(\mathbf{x},0).
\end{equation}
Now, $P(\mathbf{x},0)$ can be easily found by solving the linear anti-diffusion equation (\ref{Pt}) with
the boundary condition (\ref{Td}). Let us first consider the one-dimensional case.

\subsection{RWs in one dimension}

In one dimension, the boundary condition (\ref{Td}) becomes $P(x,T) = H(|x|-L)$.
Solving the anti-diffusion equation (\ref{Pt}) with this boundary condition, we find
\begin{equation}
\label{Psolution}
    P(x,t)=\frac{1}{2}\,\text{erfc}\left(\frac{\ell-X}{
   \sqrt{1-t/T}}\right)+\frac{1}{2}\,\text{erfc}\left(\frac{\ell+X}{\sqrt{1-t/T}}\right).
\end{equation}
At $t=0$ this yields
\begin{equation}
\label{P(x,0)}
    P(x,0)=\frac{1}{2}\,\text{erfc}\left(\ell-X\right) +\frac{1}{2}\,\text{erfc}\left(\ell+X\right).
\end{equation}
Plugging it into Eq.~(\ref{SRW1}), we arrive at
\begin{equation}
\label{Squ}
-\ln {\cal P}(L, T,n)  \simeq S=n \sqrt{4T} \mathfrak{s} (\ell),
\end{equation}
where
\begin{equation}
\label{P1_int1}
\mathfrak{s}(\ell) = \int_{-\infty}^\infty dX\, \ln\frac{2}{\text{erfc}(\ell - X)+\text{erfc}(\ell+X)}.
\end{equation}
One can see that $\ln {\cal P}(L, T,n)$ indeed exhibits dynamic scaling, as expected for all diffusive gases, see Sec. \ref{scaling}. The linear dependence on the density, $S\propto n$, reflects the non-interacting character of RWs.

Let us find asymptotic behaviors of $\mathfrak{s}(\ell)$. For $\ell\ll 1$, Eq.~(\ref{P1_int1}) yields
\begin{eqnarray}
\label{P1_small}
\mathfrak{s}(\ell) &=& 2\ell +\sqrt{\frac{2}{\pi}}\,\ell^2+ \ldots,
\end{eqnarray}
so that
\begin{equation}\label{P1_small1}
  \ln \mathcal{P}(L, T,n )  \simeq - n \left(2L+ \frac{L^2}{\sqrt{2\pi T}}+\dots\right).
\end{equation}
We emphasize two important features. First, the leading-order term, $\ln \mathcal{P}(L,T)\sim -2 n L$, coincides with the \emph{annealed} result for the RWs, see Table 2 with $V=2 L$. This is anticipated as in the limit of $\ell \to 0$ (small $L$ or large $T$), the system is given sufficient time to approach equilibrium and exploit  equilibrium fluctuation for creating an optimal initial density profile which facilitates the void formation. We expect this feature to hold for \emph{interacting} diffusive gases and confirm this expectation numerically for the SSEP in Sec. \ref{num}. Second, the void formation probability in the quenched setting is (exponentially) smaller than the void formation probability in the annealed setting. This feature is general: by fixing $q(\mathbf{x},0)=n$, we narrow the class of density variations in the problem of maximizing ${\cal P}$. As a result, ${\cal P}_{\text{quenched}}\leq {\cal P}_{\text{annealed}}$, see also Appendix A2.

\begin{figure}
\includegraphics[width=2.5 in,clip=]{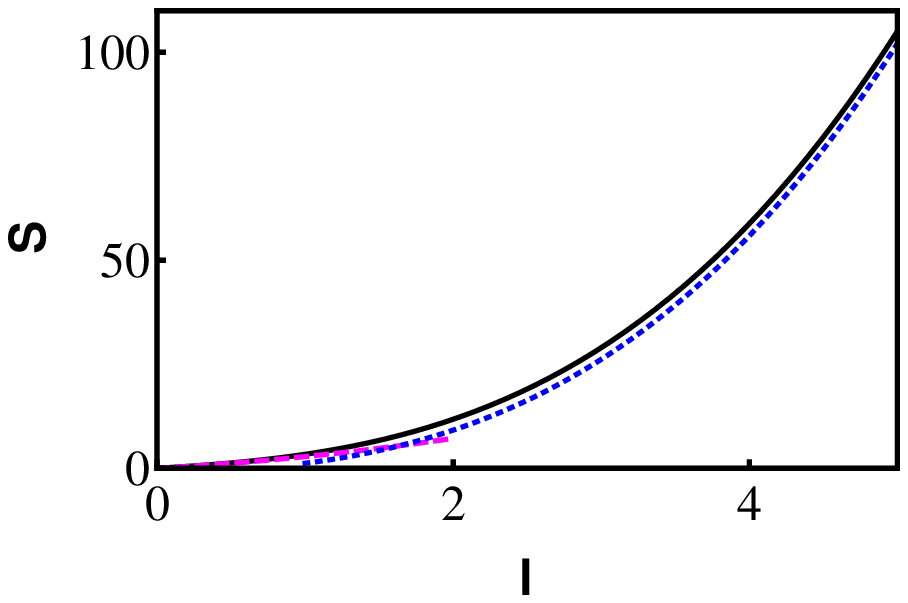}
\caption{(Color online)} Function $\mathfrak{s}(\ell)$ from Eq.~(\ref{P1_int1}) (solid line),
the small-$\ell$ asymptotic (\ref{P1_small}) (dashed line) and the large-$\ell$ asymptotic (\ref{P1_better}) (dotted line).
\label{S(ell)}
\end{figure}

In the far-from-equilibrium limit,  $\ell \gg 1$, we can employ the large-$y$ asymptotic of $\text{erfc}(y)$,
\begin{equation}
\label{erfc_asymp}
\text{erfc}(y)
=\frac{e^{-y^2}}{\sqrt{\pi}}\left(\frac{1}{y}-\frac{1}{2y^3}+\ldots\right),
\end{equation}
to extract the asymptotic of the integral in \eqref{P1_int1}. The main contribution to the integral is gathered in the region $|Z|<\ell$, and we obtain
\begin{equation}
\label{largel}
\mathfrak{s}(\ell) = \frac{2}{3}\, \ell^3 +2\ell \ln\ell + [\ln(4\pi) -2]\ell + \mathcal{O}(1).
\end{equation}
This leads to
\begin{equation}
\label{P1_better}
\ln \mathcal{P}(L, T,n)  \simeq -n\left[\frac{L^3}{6T} + 2 L\ln\!\left(\frac{L}{e}\sqrt{\frac{\pi}{T}}\right)\right].
\end{equation}
Note that the leading term of the asymptotic (\ref{largel}), $(2/3)\, \ell^3$ comes from approximating the integral in Eq.~(\ref{P1_int1}) as  $\int_{-\ell}^{\ell} dz (z-\ell)^2$. This is the integral of the squared distance from the boundary of the rescaled void over the rescaled void. As we will see shortly, this geometric property holds, at $\ell\gg 1$, for RWs in any dimension. Figure \ref{S(ell)} depicts  $\mathfrak{s}(\ell)$ along with its asymptotic behaviors. One can see that it is exponentially less probable that a void of a given size appears in a short time, than in a long time.

\begin{figure}
\includegraphics[width=2.5 in,clip=]{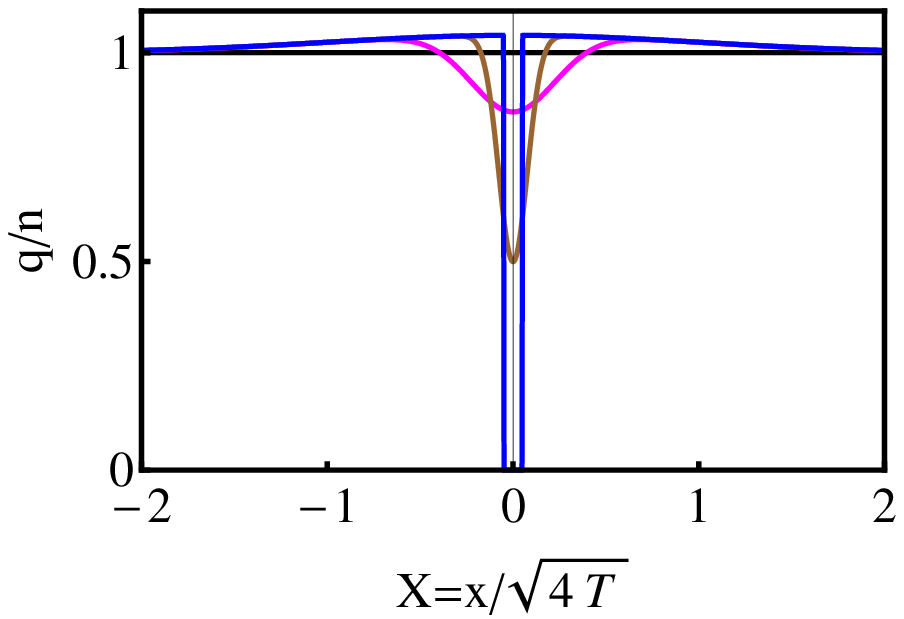}
\includegraphics[width=2.5 in,clip=]{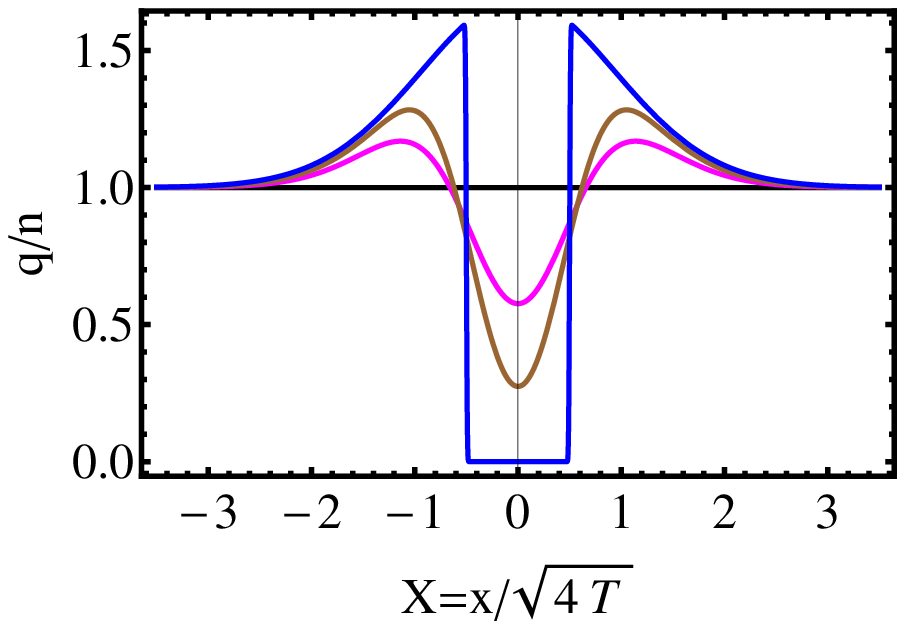}
\includegraphics[width=2.5 in,clip=]{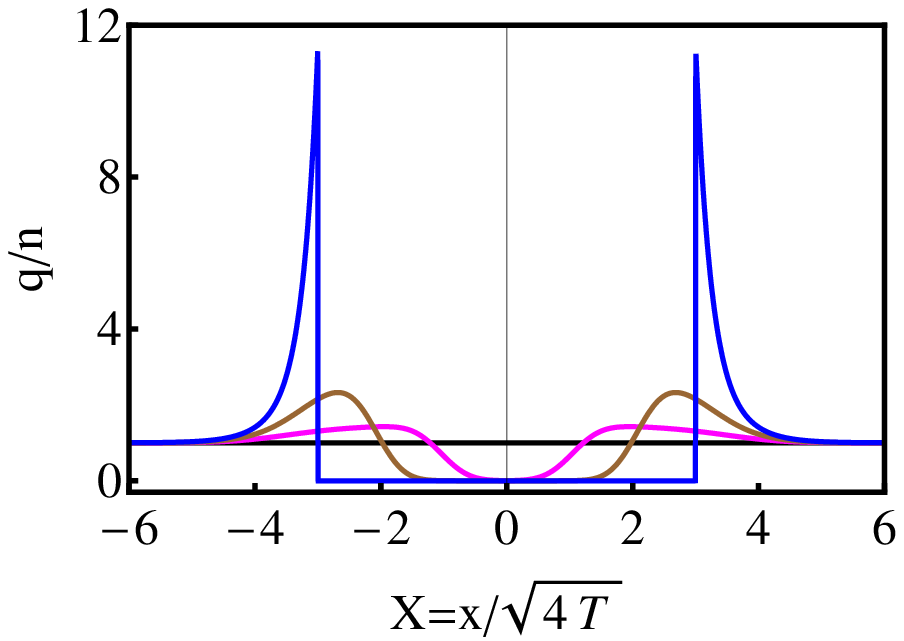}
\caption{(Color online)} Void formation in the quenched case for independent random walkers. Shown are the optimal density histories $q(X,t)$ at times $t/T=0, 0.9, 0.99$ and $1$ for  $\ell=0.05$ (left panel), at times $t/T=0, 0.5, 0.75$ and $1$ for $\ell=0.5$ (right panel), and at times $t/T=0, 1/3, 2/3$ and $1$ for $\ell=3$ (bottom panel).
\label{quvoid}
\end{figure}

Now let us see how the optimal density field evolves in time.  Using Eq.~(\ref{P(x,0)}) and the condition $q(x,0)=n$, we obtain
\begin{equation}
\label{Q(x,0)}
 Q(x,0)=\frac{n}{P(x,0)}=\frac{2 n} {\text{erfc}\left(\ell-X\right)+\text{erfc}\left(\ell+X\right)}
\end{equation}
Solving the diffusion equation (\ref{Qt}) with this boundary condition yields
\begin{equation}\label{diff}
    Q(x,0< t\leq T)=n\,\sqrt{\frac{4 T}{\pi t}}\int_{-\infty}^{\infty} \frac{e^{-\frac{T}{t}\,(X-z)^2}\,dz}
    {\text{erfc}\left(\ell-z\right)+\text{erfc}\left(\ell+z\right)}.
\end{equation}
Equations (\ref{Psolution}) and (\ref{diff}) together with relation  $q=QP$ give us the optimal density.
In particular,
\begin{equation}
\label{t=T}
q(x,T)= \frac{2n\,H(|x|-L)}{\sqrt{\pi}}\int_{-\infty}^{\infty} \frac{e^{-(X-z)^2}\,dz}{\text{erfc}\left(\ell-z\right)+\text{erfc}\left(\ell+z\right)}.
\end{equation}
The optimal density history in the quenched case does not coincide with the time-reversed solution of the diffusion equation. This can be seen, for different values of $\ell$, from Fig. \ref{quvoid}. For $\ell \gtrsim 1$ there is a striking difference between the annealed and quenched density histories
outside the void. For $l\ll 1$ (the left panel), the annealed and quenched optimal density histories become similar. In particular,  most of the void formation in this case occurs towards the end of the time interval $0<t<T$, when the system can exploit (almost) equilibrium fluctuations.

As one can see, the gas density profile outside of the quenched void at $t=T$ is quite non-trivial. For instance, there are density cusps at the void boundaries $x=\pm L$. The magnitude of the density at the peaks is
\begin{equation}
q(\pm \,L,T)=\frac{2 n}{\sqrt{\pi}}\int_{-\infty}^\infty dz\,\,
\frac{e^{-(\ell-z)^2}}{\text{erfc}(\ell-z)+\text{erfc}(\ell+z)}.
\end{equation}
For $\ell \ll 1$ Eq.~(\ref{t=T}) becomes
\begin{equation*}
q(|x|>L,T)\simeq n \left(1+\sqrt{\frac{2}{\pi}}\, \ell \,e^{-X^2/2} \right)
\end{equation*}
which is close to $n$ everywhere.  On the contrary, for $\ell \gg 1$ the gas density at the  boundaries is much greater than $n$:
\begin{equation}
\label{density_front}
q(\pm L,T) \simeq n\left[\ell^2 +\ln\ell + \tfrac{1}{2}\ln(4\pi) + {\cal O}(\ell^{-2})\right].
\end{equation}
It rapidly decays, however, as a function of $x$, and approaches the initial density $n$. The decay occurs in a boundary layer of width $\ell^{-1}\ll 1$. Indeed, writing $|X|-\ell = \xi/\ell$ and taking the limit $\ell\to\infty$ with $\xi$ kept finite, we can simplify the integral in Eq.~\eqref{t=T} and obtain
\begin{equation}
\label{density_B_layer}
q(x,T) =n\, \ell^2 \nu(\xi)\,, \quad \nu(\xi) =\frac{1-e^{-2\xi}-2\xi e^{-2\xi}}{2\xi^2}.
\end{equation}
In the region of $1\ll \xi\ll \ell$, or equivalently $\ell^{-1}\ll |X|-\ell \ll 1$, the gas density exhibits a power-law decay in $x$:
\begin{equation}
\label{density_out}
q(x,T) \simeq \frac{n}{2(|X|-\ell)^2} = \frac{2 n T}{(|x|-L)^2}.
\end{equation}

Equations~(\ref{Squ}), (\ref{P1_int1}) and (\ref{t=T}) can be also obtained from exact results of the  microscopic theory, see Appendix B.

\subsection{RWs in higher dimensions}

Now let us return to Eq.~(\ref{SRW1}) and consider formation of a void of any simply-connected shape
in  $d$ dimensions. To calculate $\ln {\cal P}$, we need to solve the anti-diffusion equation (\ref{Pt}) backward in time with the boundary condition (\ref{Td}), evaluate the result at $t=0$ and plug it into Eq.~(\ref{SRW1}). In this way we obtain
\begin{equation}\label{Pd_asymp}
 \ln {\cal P}\simeq -n (4T)^{d/2} \mathfrak{s}_d,
\end{equation}
where
\begin{equation}\label{Pd}
\mathfrak{s}_d = -\int d \mathbf{Z} \ln\left[1-\pi^{-d/2} \int_{\cal V} d \mathbf{Y} e^{-(\mathbf{Z}-\mathbf{Y})^2}\right].
\end{equation}
The integration over $\mathbf{Z}$ is performed over the whole $d$-dimensional space, whereas the integration over $\mathbf{Y}$ is performed over region $\mathcal{V}$ which is obtained by rescaling all coordinates of the physical void by $\sqrt{4T}$.

\subsubsection{Small voids}

Consider voids with all characteristic dimensions much less than the diffusion length $\sqrt{4T}$. Such voids have a small rescaled volume, $|\mathcal{V}|\ll 1$. Expanding the logarithm in \eqref{Pd} and keeping the leading term we get
\begin{equation}
\label{expandinglog}
   \mathfrak{s}_d(\mathcal{V})
\simeq \int d{\bf Z}\,\pi^{-d/2}\int_\mathcal{V} d{\bf Y}\,e^{-({\bf Z}-{\bf Y})^2}\,.
\end{equation}
Exchanging the order of integration we find that the integral reduces to $|\mathcal{V}|$.
This leads to
\begin{equation}
\label{equilibrium}
\mathcal{P}_d(V, T)\simeq e^{-n V},
\end{equation}
which coincides with the annealed void formation probability for RWs, see Table II.  Expanding the logarithm in \eqref{Pd} to higher orders, we find
\begin{equation}
\label{small_void}
\mathfrak{s}_d(\mathcal{V})  = |\mathcal{V}| + \tfrac{1}{2}(2\pi)^{-d/2} |\mathcal{V}|^2 +\ldots
\end{equation}
Surprisingly, the sub-leading term also depends only on the rescaled volume $|\mathcal{V}|$ and is independent of the shape of the void. In one dimension $|\mathcal{V}|=2 \ell$, and Eq.~(\ref{small_void})
coincides with Eq.~(\ref{P1_small}).

The equilibrium result (\ref{equilibrium}) actually holds under a much weaker assumption on the spatial dimensions of the rescaled void. It suffices to demand the following strong inequality:
\begin{equation}
\label{heating}
\max \limits_{\mathbf{X}} \mathcal{D}({\bf X})\ll 1,
\end{equation}
where $\mathcal{D}({\bf X})$ is the shortest distance from ${\bf X}$ to the boundary $\partial \mathcal{V}$ of the rescaled void. Under condition (\ref{heating}), the field $P(\mathbf{X},0)$
that determines the action (\ref{SRW1}) is close to unity in all points inside the void, validating
the approximation (\ref{expandinglog}) leading to Eq.~(\ref{equilibrium}).

\subsubsection{Large voids}

Now consider voids  with all characteristic dimensions much greater than the diffusion length $\sqrt{4T}$.
Such voids have a large rescaled volume, $|\mathcal{V}|\gg 1$. In this case it is convenient to rewrite \eqref{Pd} as
\begin{equation}
\label{Pd_2}
\mathfrak{s}_d(\mathcal{V}) = -\int d{\bf Z}\,
\ln\left[\pi^{-d/2}\int_{\overline{\mathcal{V}}} d{\bf Y}\,e^{-({\bf Z}-{\bf Y})^2}\right],
\end{equation}
where the second integration now goes over the exterior $\overline{\mathcal{V}}=\mathbb{R}^d-\mathcal{V}$ of the rescaled void $\mathcal{V}$. The term inside the logarithm in Eq.~\eqref{Pd_2} is very close to 1 when ${\bf Z}\in \overline{\mathcal{V}}$, apart from the region very close to the boundary of the rescaled void $\mathcal{V}$, but the contribution from this region is very small as the rescaled void is large. For the points ${\bf Z}\in \mathcal{V}$, the term inside the logarithm is very small. This provides the dominant contribution to the first integral in \eqref{Pd_2} and implies a simplification. Indeed, let ${\bf Y}_{\bf M}$ be the point on the boundary $\partial \mathcal{V}$ of the rescaled void $\mathcal{V}$ which is closest to ${\bf Z}$. In other words, $\mathcal{D}({\bf Z}) = |{\bf Z}-{\bf Y}_{\bf Z}|$ is the shortest distance from ${\bf Z}$ to the rescaled void boundary $\partial \mathcal{V}$. Then a crude estimate of the internal integral in Eq.~(\ref{Pd_2}),
\begin{equation}
\pi^{-d/2}\int_{\overline{\mathcal{V}}} d{\bf Y}\,e^{-({\bf Z}-{\bf Y})^2}\sim
e^{-[\mathcal{D}({\bf Z})]^2}
\end{equation}
suffices to provide a correct leading contribution to Eq.~\eqref{Pd_2}. Thus we arrive at
\begin{equation}
\label{large_void}
\mathfrak{s}_d(\mathcal{V})  \simeq \int_\mathcal{V} d{\bf Z}\,\, [\mathcal{D}({\bf Z})]^2,
\end{equation}
a purely geometric result.

Now we are in a position to ask the following question.  Given that a void of volume $V$ has formed in a diffusive lattice gas at time $T$, what is the most likely void shape? To get insight, let us first apply Eq.~(\ref{large_void}) to a $d$-dimensional spherical void of rescaled radius $\ell=R/\sqrt{4T} \gg 1$. We obtain
\begin{equation}
\label{Pd_large_ball}
\mathfrak{s}_d^{\text{ball}} \simeq d v_d\int_0^\ell dr\,r^{d-1}\,(\ell-r)^2 = \frac{2v_d \ell^{d+2}}{(d+1)(d+2)} =\frac{2  \,v_d^{-2/d} |\mathcal{V}|^{1+\frac{2}{d}}}{(d+1)(d+2)},
\end{equation}
where $v_d=\pi^{d/2}/\Gamma(1+d/2)$ is the volume of the unit sphere in $d$ dimensions. Combining Eqs.~\eqref{Pd_asymp} and \eqref{Pd_large_ball} and recalling that $|\mathcal{V}|= v_d \ell^{d}$ we obtain
\begin{equation}
\label{Pd_V}
\mathcal{P}_d(V, T,n)  \sim \exp\!\left[-\frac{n v_d^{-\frac{2}{d}}}{(d+1)(d+2)}\frac{V^{1+\frac{2}{d}}}{2T}\right].
\end{equation}
Similarly,  for a cube-shaped void we find
\begin{equation}
\label{Pd_large_cube}
\mathfrak{s}_d^{\text{cube}} \simeq \frac{|\mathcal{V}|^{1+\frac{2}{d}}}{2(d+1)(d+2)}.
\end{equation}
An inspection shows that, for a given $|\mathcal{V}|$, and for $d>1$, $\mathfrak{s}_d^{\text{cube}}<\mathfrak{s}_d^{\text{ball}}$. That is, at short times, the formation of a spherical void is (exponentially) less probable than the formation of a cube-shaped void of the same volume.  But the cube is also far from the most probable shape at a given void volume. This becomes clear when one
uses Eq.~\eqref{large_void} to calculate $\mathfrak{s}_d$ for a $d$-dimensional rectangular parallelepiped (cuboid). For example, for a two-dimensional rectangle with rescaled sides of lengths $a\gg 1$ and $b\gg 1$, $a\geq b$, we obtain
\begin{equation}
\label{P_rectangle}
\mathfrak{s}_2^{\text{rectangle}} = \frac{b^3(2a-b)}{24} = \frac{|\mathcal{V}|^2}{24}\,\frac{2\kappa-1}{\kappa^2},
\end{equation}
where $\kappa=a/b\geq 1$ is the aspect ratio of the rectangle, and $|\mathcal{V}|=ab$ is the rectangle area. The function $\kappa^{-2} (2\kappa-1)$, entering Eq.~(\ref{P_rectangle}),
has its maximum at $\kappa=1$ (that is, for a square-shaped void) and is monotone decreasing at $\kappa>1$. Hence the minimum of $\mathfrak{s}_d$ at fixed $|\mathcal{V}|$ is achieved at very large $\kappa$ (formally at $\kappa \to \infty$), when the rectangle becomes a long and thin filament. For thin filaments, however, the large-void approximation, Eq.~(\ref{large_void}), breaks down. On the other hand, in the limit of $a\gg 1$ and $b\ll 1$ the equilibrium asymptotic
(\ref{small_void}) holds, which is \emph{independent}, in the leading and sub-leading order, of the void shape.  Generalizing to $d$ dimensions, we see that the most likely void shape, at a given volume, is such that at least one characteristic dimension is much smaller than the diffusion length $\sqrt{4T}$. In this case the void formation probability, in the leading order, is close to the equilibrium probability which is independent of the shape. At this level of accuracy we cannot distinguish between a whole variety of shapes: for example, between a pancake and a filament of the same volume in $3d$.  To determine the most likely shape, one may need to account for non-equilibrium sub-leading terms in $\mathfrak{s}_d$ that can still be important, as they appear in the exponent.

\subsubsection{Spherical voids}

Among voids of a given large volume, spherical voids arise with the lowest probability; this assertion is rather evident, although we do not have a rigorous proof.  Let $\mathcal{V}$ be the rescaled ball of rescaled radius $\ell$. Writing $|{\bf Z}|=r$ and using spherical symmetry we get $d{\bf Z}= d v_d r^{d-1} dr$. Simplifying the integral in \eqref{Pd}, we obtain
\begin{equation}
\label{Pd_integral}
\mathfrak{s}_d(\ell) = -d v_d\int_0^\infty dr\,r^{d-1}\,\ln f_d(r,\ell)
\end{equation}
where
\begin{equation*}
f_d(r,\ell) =1-\pi^{-d/2}e^{-r^2}\int_{|{\bf Y}|<\ell} d{\bf Y}\,e^{2{\bf Z}\cdot {\bf Y} -|{\bf Y}|^2}.
\end{equation*}
In three dimensions, for instance,
\begin{equation}
\label{P3_integral}
\mathfrak{s}_3(\ell) = - 4\pi\int_0^\infty dr\,r^2\,\ln f_3(r,\ell),
\end{equation}
and $f_3(r,\ell)$ simplifies to
\begin{equation}
\label{F3_integral}
f_3(r,\ell) = \frac{\text{erfc}(\ell - r)+\text{erfc}(\ell + r)}{2}
+ \frac{e^{-(\ell - r)^2}-e^{-(\ell + r)^2}}{2\sqrt{\pi}\,r}.
\end{equation}
Now we can return to the $\ell\to \infty$ limit and calculate a sub-leading correction to the already known leading asymptotic \eqref{Pd_large_ball}. Using Eq.~\eqref{erfc_asymp}, we obtain
\begin{equation}
\mathfrak{s}_3(\ell) = \frac{2\pi}{15}\,\ell^5 +
\frac{4\pi}{3}\,\ell^3\left[\ln\ell +\frac{1}{2}\,\ln(4\pi)-\frac{13}{6}\right]+ \dots .
\end{equation}

\section{Void formation in quenched setting: SSEP}
\label{num}

Now we briefly consider the void formation in the SSEP. For the quenched setting, Eqs.~(\ref{d1}) and (\ref{d2}) cannot be solved analytically. We solved these equations numerically in one spatial dimension, by using an iteration algorithm  originally developed by Chernykh and Stepanov \cite{Stepanov} for evaluating the probability distribution of large negative velocity gradients in the Burgers turbulence. Different modifications of this algorithm have been used for evaluating large deviation functions of several lattice gas models, with and without on-site reactions \cite{EK,MS2011,KM_var}. This algorithm  is ideally suitable for the void formation problem, as this problem involves mixed boundary conditions in time: one condition for $q$, $q(x,0)=n$,  and another for $p$, see Eqs.~(\ref{pT}) and (\ref{pT2}). The algorithm iterates the diffusion-type Eq.~(\ref{d1}) forward in time and the anti-diffusion-type Eq.~(\ref{d2}) backward in time (see \cite{KM_var} for details).  To suppress numerical instability, we used a linear combination of values of $q$ and $p$ from \emph{two} previous iterations when solving for $p$ and $q$, respectively \cite{Stepanov,KM_var}. The only significant difference between our present implementation of the algorithm and that of Ref. \cite{KM_var} comes from the boundary condition $p(|x|<L,T)=-\infty$ which is inconvenient numerically. To overcome this inconvenience, we performed the Hopf-Cole canonical transformation from $q$ and $p$ to $Q=q e^{-p}$ and $P=e^p$.
The transformed Hamiltonian density is
\begin{equation}
\label{W1}
    \widetilde{\mathcal{H}}=-\partial_x Q \,\partial_x P-Q^2(\partial_x P)^2,
\end{equation}
and the new Hamilton equations read
\begin{eqnarray}
  \partial_t Q &=& \partial_{xx} Q+2\partial_x (Q^2\partial_x P), \label{QP1a} \\
  \partial_t P &=& - \partial_{xx} P+2Q(\partial_x P)^2. \label{QP1b}
\end{eqnarray}
As a result,  the action becomes
\begin{eqnarray}
\label{ActionSSEP}
S &=& \int\int \left(P\partial_t Q-\widetilde{\mathcal{H}}\right)\, dx\, dt+\int \Phi(q,P)\, dx\biggl|_0^T\nonumber\\
&=&
- \int \int Q^2 (\partial_x P)^2 \,dx\,dt-n \int dx \ln P(x,0),
\end{eqnarray}
where $\Phi(q,P)=q \ln P$, see Sec. \ref{RW} A.  The boundary conditions in time become $Q(x,0) P(x,0)=n$ and $P(x,T)=H(|x|-L)-1$. The boundary conditions
in $x$ do not change: $Q(|x|\to \infty,t)=1$ and $P(|x|\to \infty,t)=0$.

\begin{figure}[ht]
\includegraphics[width=2.4 in,clip=]{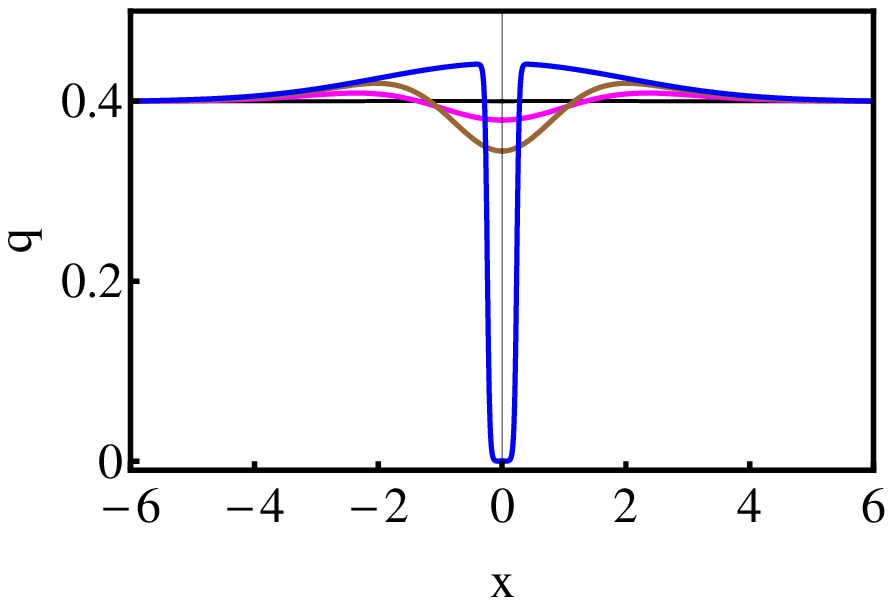}
\includegraphics[width=2.4 in,clip=]{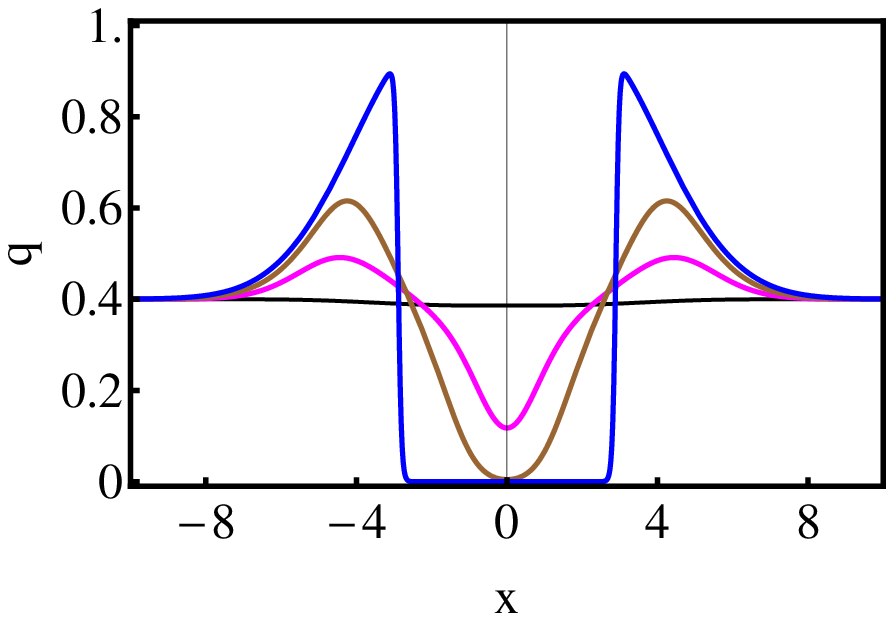}
\caption{(Color online) Void formation in the quenched case for the SSEP. Shown are numerically computed optimal density histories $q(x,t)$  for $L=0.25$ and $T=1$ (left panel) and for $L=3$ and $T=1$ (right panel) at time moments $t=0$, $1/3$, $2/3$ and $1$. The initial density $n=0.4$}
\label{SSEPq}
\end{figure}

We implemented this algorithm in {\it Mathematica}, working with finite systems of reasonable sizes. The step-function entering the boundary conditions for $P$ at $t=T$ was smoothed a bit. The iterations converge rapidly. Mass conservation in the numerical box was used to monitor the accuracy. Having computed $Q(x,t)$ and $P(x,t)$, we determined $q(x,t)=Q(x,t) P(x,t)$ and computed the action $S$ by numerically evaluating the integrals in Eq.~(\ref{ActionSSEP}).

Figure \ref{SSEPq} shows two examples  of numerically found optimal density histories $q(x,t)$ for $n=0.4$ and $T=1$: for a small void, $L=0.25$, that is $\ell =0.125$ (the left panel) and a relatively large void, $L=3$, that is $\ell=1.5$. One can see that for $L=0.25$ the density history is similar to those for RWs with small $\ell$, as shown in the two upper panels of Fig. \ref{quvoid}. For large $\ell$ this similarity must break down, as  the gas density in the SSEP cannot exceed $q=1$.  Furthermore, for very large voids, $\ell \gg 1$, the characteristic decay length of the density outside the void must behave as $\ell$, the same as the displaced mass. (This should be contrasted with the RWs where, at $n=1$, the maximum gas density is $\ell^2\gg 1$, and the density decay length is $1/\ell \ll 1$, see Sec. \ref{RW} B.)
These two features of the SSEP can be discerned already at $\ell=1.5$, see the right panel of Fig. \ref{SSEPq}.  We could not probe $\ell>1.5$, as the numerical accuracy became insufficient. 

\begin{figure}[ht]
\includegraphics[width=2.6 in,clip=]{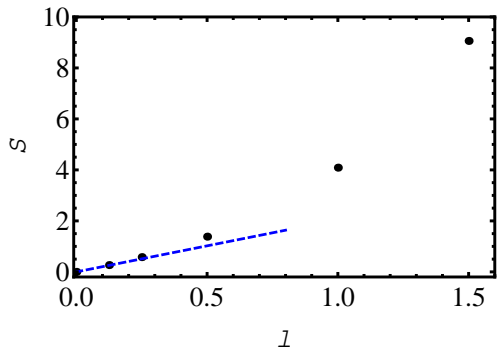}
\caption{(Color online). $\mathcal{S}(\ell,n)$ versus $\ell$ for the quenched setting of the SSEP with
$n=0.4$. Circles: numerical results obtained with the iteration algorithm. Dashed line: the
annealed result $\mathcal{S}=4\ell \, \ln[1/(1-n)]$, see the text.}
\label{S_SSEP}
\end{figure}

Figure~\ref{S_SSEP} depicts the $\ell$-dependence of the rescaled action  $\mathcal{S}(\ell,n)$ that we found numerically in a moderate range of $\ell =L/\sqrt{4T}$ for $n=0.4$. As expected, at small $\ell$ the rescaled action approaches that for the annealed case, $\mathcal{S}_{\text{annealed}} = 4\ell \ln[1/(1-n)]$. The latter relation follows from Eq.~(\ref{annealedP}), see Table II, and the dynamic scaling relation $S(L,T,n)=T^{1/2} \mathcal{S}(\ell,n)$, see Sec. \ref{scaling}. As also expected, the rescaled action for the quenched setting is greater than the one for the annealed setting, so that ${\cal P}_{\text{quenched}} < {\cal P}_{\text{annealed}}$.

\section{Discussion}
\label{summary}

We investigated the probability of macroscopic void formation in a class of diffusive lattice gases whose hydrodynamic description is provided by the diffusion equation (\ref{rho:eq}). The formalism of macroscopic fluctuation theory is perfectly suitable for the analysis of
the void formation problem --- it predicts the dynamic scaling behavior (\ref{scalingform}), and it also yields the most likely density history of the system in the process of void formation. In the annealed setting the void formation probability turns out to be independent of $T$ and given by the Gibbs-Boltzmann equilibrium formula. The quenched setting is harder to study, and we were only able to solve analytically the case of  non-interacting random walkers. Fortunately, a relatively straightforward numerical treatment is feasible for interacting particles, as we have demonstrated for the simple symmetric exclusion process. An interesting avenue for the future research is to develop an analytical theory in the limit of large voids. The hope is to use asymptotic methods to circumvent the challenge of (most likely, inaccessible) exact solution of the MFT equations for interacting lattice gases. More specifically, it would be interesting to find whether the super-Gaussian tail $\ln \mathcal{P} \simeq - f(n) L^3/T$, in analogy with our RW result (\ref{P1_better}), is universal for a whole class of interacting lattice gases. A similar question  about the super-Gaussian tail $\ln \mathcal{P}(J) \sim -J^3/T$ of the probability to observe a very large integrated current $J$ in one dimension was raised in \cite{DG2009b}, and the arguments in favor of this behavior were given in Refs.~\cite{DG2009b,varadhan}.

We argued that the probability of formation of macroscopic voids is a meaningful way to characterize large deviations in classical diffusive lattice gases. As we already mentioned in the Introduction, a similar
characterization of large deviations has been employed for quantum many-body systems, see \textit{e.g.} \cite{Abanov1,Abanov2}, where it is known under the name of ``emptiness formation".  One approach to finding the emptiness formation probability relies on exact solutions of integrable models \cite{Korepin}, and subsequent computation of the large-distance asymptotic behaviors, see \cite{Abanov2}. Another approach is via an effective nonlinear hydrodynamic description that directly probes the large-distance asymptotics \cite{Abanov1,Abanov2}. In the latter description the probability of void formation corresponds to an activation trajectory: an instanton solution of a classical hydrodynamics (or rather ``anti-hydrodynamics" where, in $1d$,   the usual pair of acoustic waves in a compressible gas give way to a pair of aperiodic modes: one growing, the other decaying). On a qualitative level, this ``anti-hydrodynamics" may be compared to the ``anti-diffusion" that appears in the lattice gas settings we dealt with here. This analogy opens new exciting directions for future work.

\section*{Acknowledgments}
We acknowledge a useful discussion with Omri Gat and Alex Kamenev.
B.M. was supported by the Israel Science Foundation (Grant
No. 408/08), by the US-Israel Binational Science Foundation
(Grant No. 2008075), and by the Condensed Matter
Theory Visitors Program of Boston University's Physics
Department. P.V.S. was supported by  the Russian Foundation for Basic Research, grant No 10-01-00463.

\appendix
\section{Derivation of the MFT equations for void formation}

Here we derive the MFT equations and boundary conditions for void formation in both annealed and quenched setting.  Rather than immediately focusing on the void formation, it proves useful to start with a more general problem and to demand $q(\mathbf{x},T)=\kappa(\mathbf{x})$ in a bounded domain $\Omega$; the void problem is then obtained by identifying $\Omega$ with the void and setting $\kappa(\mathbf{x})=0$.

For the diffusive lattice gases, the Langevin equation is \cite{Spohn}
\begin{equation}
\label{P010}
     \partial_t q =\nabla\cdot \left[D(q) \, \nabla q+\sqrt{\sigma(q)} \, \boldsymbol{\xi}(\mathbf{x},t)\right].
\end{equation}
Here $\boldsymbol{\xi}(x,t)$ is a zero-average Gaussian noise, which is delta-correlated in space and in time,
\begin{equation}
\left\langle \xi_i(\mathbf{x},t)\xi_j(\mathbf{x}^{\, \prime},t^{\prime})\right\rangle=\delta_{ij}\,
\delta(\mathbf{x}-\mathbf{x}^{\, \prime})\, \delta(t-t^{\prime})\,,
\label{P020}
\end{equation}
and the brackets denote ensemble averaging. We assume that $q(\left|\mathbf{x}\right|\to\infty,0)=n=\text{const}$. It proves useful to introduce the particle displacement $\mathbf{u}$ defined via
\begin{equation}
\label{P021}
q(\mathbf{x},t)=n+\nabla\cdot \mathbf{u} (\mathbf{x},t)
\end{equation}
and satisfying the boundary condition
\begin{equation} \mathbf{u}(\left|\mathbf{x}\right|\to\infty,0)=0 .
\label{P023}
\end{equation}
Combining (\ref{P010}) and (\ref{P021}), we deduce a Langevin equation for $\mathbf{u}$:
\begin{equation}\label{P030}
     \partial_t\mathbf{u} - D(q)\, \nabla q=\sqrt{\sigma(q)} \,\boldsymbol{\xi}\,.
\end{equation}
The probability of observing a large deviation of the displacement $\mathbf{u}(\mathbf{x},t)$ is, up to a pre-exponent,
\begin{equation}\label{pld}
{\cal P}[\mathbf{u}(x,t)]\sim \exp\!\left\{-\int_0^T dt\, \int d\mathbf{x}\,
\frac{\left[\partial_t \mathbf{u} -D(q)\nabla q\right]^2}{2\sigma(q)}\right\}.
\end{equation}
Here and in the following we assume that $q$ is related to $\mathbf{u}$ according to Eq.~(\ref{P021}).

\subsection{Annealed setting}

Let us first consider the annealed setting. At $t=0$, one starts from an (a priori unknown) optimal initial density profile $q(\mathbf{x},0)$: a certain realization of equilibrium fluctuations of the gas with average density $n=\text{const}$. The probability to observe this realization is given by the Boltzmann-Gibbs distribution and can be expressed via the function $\mathcal{F}(q)$ defined in Eq.~(\ref{modifiedF}):
\begin{equation*}
{\cal P}[q(\mathbf{x},0)]\sim \exp\!\left\{-\int d\mathbf{x}\,\mathcal{F}[q(\mathbf{x},0)]\, \right\}\, ,
\end{equation*}
The joint probability to observe a large deviation of the displacement $\mathbf{u}(\mathbf{x},t)$ when starting, at $t=0$, from $q(\mathbf{x},0)$ is
\begin{equation*}
{\cal P}[\mathbf{u}(\mathbf{x},t); q(\mathbf{x},0)]\sim \exp(-\bar{S}),
\end{equation*}
where
\begin{equation}
\label{barS}
\bar{S} = \int_0^T dt\, \int d\mathbf{x}\,
\frac{\left[\partial_t \mathbf{u} -D(q)\, \nabla q\right]^2}{2\sigma(q)}\,
+\int d\mathbf{x}\,\mathcal{F}[q(\mathbf{x},0)].
\end{equation}
Our task is to minimize $\bar{S}$ under condition that $q(\mathbf{x},T)=\kappa(\mathbf{x})$  in a bounded domain $\Omega\subset \mathbb{R}^d$. Let us define a scalar field  $p(\mathbf{x},t)$ (which will play
the role of momentum) as the unique solution of the following Poisson-type equation:
\begin{equation}
\nabla\cdot (\sigma\nabla p)=\nabla\cdot (D\, \nabla q) - \partial_t q
\label{P045}
\end{equation}
with the boundary condition
\begin{equation}
p(\left|\mathbf{x}\right|\to\infty,t)\to 0.
\label{P046}
\end{equation}
As $q$ tends to $n=\text{const}$ at $\mathbf{x}\to \infty$, the right-hand-side of Eq.~(\ref{P045}) (the ``charge density" of the Poisson equation) vanishes at $\mathbf{x}\to \infty$, and so the solution for $p$ is indeed unique. Equation~(\ref{P045})
coincides with the first of the MFT equations, Eq.~(\ref{d1}).

An immediate corollary from Eqs.~(\ref{P045}) and (\ref{P021}) is the relation
\begin{equation}
\partial_t \mathbf{u}-D\, \nabla q=-\sigma\, \nabla p +\boldsymbol{\omega}\, ,
\label{P047}
\end{equation}
where $\boldsymbol{\omega}$ is an arbitrary solenoidal vector field, $\nabla\cdot \boldsymbol{\omega}=0$, which goes to zero sufficiently fast at $\mathbf{x}\to\infty$. Although $\boldsymbol{\omega}$ does not affect $q(\mathbf{x},t)$, it does affect the action $\bar{S}$. Indeed, upon substituting Eq.~(\ref{P047}) into  Eq.~(\ref{barS}), the integral over $\mathbf{x}$ in the first term can be written as
\begin{equation}
\label{killomega}
\int  d\mathbf{x}\, \left[\frac{\sigma \left(\nabla p\right)^2}{2}+\frac{\boldsymbol{\omega}^{2}}{2\sigma}\right]
-\int  d\mathbf{x}\,\nabla\cdot\left(p\boldsymbol{\omega}\right)\, .
\end{equation}
In an infinite medium the integral of $\nabla \cdot (p \boldsymbol{\omega})$ vanishes as, by virtue of Gauss theorem, it can be reduced to a surface integral over a sphere with radius tending to infinity \cite{vortices}. Therefore, if $q(\mathbf{x},t)$ is known, the minimum of $\bar{S}$ is achieved at $\boldsymbol{\omega}=0$, and with $p$ defined as explained above. As a result, Eq.~(\ref{killomega})
yields the second equality in Eq.~(\ref{action}).

Now we calculate the variation of $\bar{S}$:
\begin{eqnarray}
\label{variation}
\delta \bar{S}&=& \int_0^T dt \int d\mathbf{x}\Biggl\{\sigma^{-1}\left[\partial_t \mathbf{u} -D\, \nabla \left(\nabla\cdot\mathbf{u}\right)\right]\, \cdot\left[\partial_t \delta \mathbf{u}-D\, \nabla \left(\nabla\cdot\delta \mathbf{u}\right)
-D^{\prime}\, \left(\nabla \left(\nabla\cdot\mathbf{u}\right)\right)\, \left(\nabla\cdot\delta \mathbf{u}\right)\right] \nonumber\\
&-&\frac{\sigma^{\prime}}{2\sigma^{2} }\left[\partial_t\mathbf{u} -D\, \nabla \left(\nabla\cdot\mathbf{u}\right)\right]^2
\left(\nabla\cdot\delta \mathbf{u}\right) \Biggr\}
+\int d\mathbf{x}\,\mathcal{F}^{\prime}(q(\mathbf{x},0))\, \delta q(\mathbf{x},0) = 0.
\end{eqnarray}
Using Eq.~(\ref{P047}) with $\boldsymbol{\omega}=0$, we can rewrite the double integral  in Eq.~(\ref{variation}) as $\int_0^T dt \int d\mathbf{x} \,\mathcal{W}$ with
\begin{equation*}
\mathcal{W}= - \nabla p \cdot \left[\partial_t \delta \mathbf{u}-D\, \nabla \left(\nabla\cdot\delta \mathbf{u}\right)
-D^{\prime}\, \left(\nabla \left(\nabla\cdot\mathbf{u}\right)\right)\, \left(\nabla\cdot\delta \mathbf{u}\right)\right]
-\frac{\sigma^{\prime}}{2} \left(\nabla p\right)^2 \left(\nabla\cdot \delta \mathbf{u}\right).
\end{equation*}
The first term, $-\nabla p \cdot \partial_t \delta \mathbf{u}$, can be integrated over time by parts. Then, grouping the terms with $D$ and $D^{\prime}$ together and employing Green's  first identity and the boundary condition $\mathbf{u}(|\mathbf{x}| \to \infty,t)=0$,
we obtain
\begin{eqnarray}
 \delta \bar{S} &=& - \int_0^T dt \, \int d\mathbf{x}\,
  \Big[\partial_t p+D\, \nabla^2 p+\frac{1}{2}\sigma^{\prime} \left(\nabla p\right)^2\Big]\, \left(\nabla\cdot\delta \mathbf{u}\right)\nonumber\\
&+& \int  d\mathbf{x}\,\delta \mathbf{u}(\mathbf{x},0)\cdot\Big[\nabla p(\mathbf{x},0)-\nabla \mathcal{F}^{\prime} (q(\mathbf{x},0))\Big]\nonumber\\
  &-& \int d\mathbf{x} \,\delta \mathbf{u}(\mathbf{x},T)\cdot\nabla p(\mathbf{x},T) = 0.
\label{P050}
\end{eqnarray}
Demanding the bulk term to vanish we get Eq.~(\ref{d2}),
the second of the MFT equations. Each of the two boundary terms in Eq.~(\ref{P050}) must vanish independently. The first term yields the boundary condition at $t=0$:
\begin{equation*}
p(\mathbf{x},0)-\mathcal{F}^{\prime}[q(\mathbf{x},0)]=A=\text{const}.
\end{equation*}
In view of the boundary conditions~(\ref{P046}) and~(\ref{P023}) and relation $\mathcal{F}^{\prime}(n)=0$, we obtain $A=0$ which yields Eq.~(\ref{annealed0}).

Now let us turn to the second boundary term in Eq.~(\ref{P050}) and assume that the domain $\Omega$ is simply connected. As it turns out, the momentum field $p(\mathbf{x},T)$ experiences a jump at the boundary $\partial \Omega$ of the domain $\Omega$. To have a minimum action the third term in Eq.~(\ref{P050}) must vanish:
\begin{equation}
\int d\mathbf{x} \,\delta \mathbf{u}(\mathbf{x},T)\cdot\nabla p(\mathbf{x},T) = 0\, .
\label{P054}
\end{equation}
This equality should be valid for an arbitrary $\delta \mathbf{u}(\mathbf{x},T)$ compatible with the boundary conditions. In particular, it must vanish for any sufficiently smooth $\delta \mathbf{u}(\mathbf{x},T)$ that vanishes at
$\mathbf{x} \in \Omega\cup \partial \Omega$. This means that  $\nabla p$ must vanish at $\mathbf{x}\in\overline{\Omega}$, where $\overline{\Omega}=\mathbb{R}^d-\Omega$ is the exterior of the domain $\Omega$. This,  along with the condition (\ref{P046}), yields
\begin{equation}
p(\mathbf{x}\in\overline{\Omega},T)=0\, .
\label{P056}
\end{equation}
Hence Eq.~(\ref{P054}) can be rewritten as
\begin{equation*}
\int\limits_{\widetilde{\Omega}} d\mathbf{x} \,\delta \mathbf{u}(\mathbf{x},T)\cdot\nabla p(\mathbf{x},T) = 0
\end{equation*}
for any $\widetilde{\Omega}$ which contains $\Omega$ and its boundary $\partial \Omega$. Integrating this equation by parts and taking into account Eq.~(\ref{P056}), we obtain:
$$
\int\limits_\Omega d\mathbf{x}\, p(\mathbf{x},T)\, \nabla\cdot\delta\mathbf{u}(\mathbf{x},T) +
\int\limits_{\mathbf{x}\in \partial \Omega}  dS\,\delta u_n(\mathbf{x},T)\,  p_{in} (\mathbf{x},T) =0\, ,
$$
where $\delta u_n = \delta\mathbf{u} \cdot \mathbf{n}$, $\mathbf{n}$ is the external unit normal to $\partial \Omega$, and
$p_{in} (\mathbf{x},T)\bigr|_{\mathbf{x}\in\partial\Omega}=
\lim_{\mathbf{x}\to\partial\Omega~(\mathbf{x}\in\Omega)} p(\mathbf{x},T)$ is the value of $p(\mathbf{x},T)$ along the boundary $\partial\Omega$ inside $\Omega$.
As
$\delta q(\mathbf{x},T)=\nabla\cdot\delta \mathbf{u}(\mathbf{x},T)=0$ inside $\Omega$, the first term in Eq.~(\ref{P054}) vanishes.
Now we need to deal with the second term:
\begin{equation}
\label{P054a}
\int\limits_{\mathbf{x}\in \partial \Omega}  dS\,\delta u_n(\mathbf{x},T)\, p_{in}(\mathbf{x},T) =0.
\end{equation}
Let us apply Gauss theorem to the following auxiliary integral:
\begin{equation}
\label{Gauss1}
\int\limits_{\mathbf{x}\in \partial \Omega}  dS\, u_n(\mathbf{x},T) =
\int\limits_{\mathbf{x}\in \Omega}d\mathbf{x}\, \nabla \cdot\mathbf{u}(\mathbf{x},T).
\end{equation}
As $q(\mathbf{x}\in\Omega,T)=\kappa(\mathbf{x})$, the value of integral in the right side of Eq.~(\ref{Gauss1}) is fixed, see Eq.~(\ref{P021}). Hence, the variation $\delta \mathbf{u}$ must obey $\int_{\mathbf{x}\in \partial \Omega}  dS\, \delta u_n(\mathbf{x},T)=0$. Comparing this result with Eq.~(\ref{P054a}) we see that $p_{in}$ must be constant along $\partial \Omega$.

Overall, the complete set of boundary conditions at $t=0$ and $t=T$ for the annealed setting reads:
\begin{equation*}
p(\mathbf{x},0)=\mathcal{F}^{\prime}[q(\mathbf{x},0)] ;
\end{equation*}
\begin{equation}
\label{BCT}
q(\mathbf{x}\in\Omega,T)=\kappa(\mathbf{x});\quad p(\mathbf{x}\in\overline{\Omega},T)=0; \quad p_{in}\big|_{t=T}=\text{const},
\end{equation}
where the constant is unknown \textit{a priori} and is a part of the solution. With these boundary conditions, the solution turns out to  be quite simple, see Sec. \ref{genannealed}, because of
the presence of a local integral of motion $p=\mathcal{F}^{\prime}(q)$. In particular, one can immediately see from this integral of motion that  $q(\mathbf{x}\in\overline{\Omega},T)=n$. Furthermore,
in the particular case of void, $q(\mathbf{x}\in\Omega,T)=0$, one obtains $p(\mathbf{x}\in\Omega,T)=-\infty$  for all lattice gases with $\mathcal{F}^{\prime}(0)=-\infty$. These include the RWs, the SSEP, the KMP, and the ZRP with $\alpha(0)=0$.

\subsection{Quenched setting}

In the quenched case $q(\mathbf{x},0)=n$ is specified, and the second term in  Eq.~(\ref{barS})
vanishes. 
To proceed, we can simply put $\mathcal{F}=0$ in Eq.~(\ref{P050}) leading to
\begin{eqnarray}
   \delta \bar{S} &=& -\int_0^T dt \, \int d\mathbf{x}\,
  \Big[p_{t}+D\, \nabla^2 p+\frac{1}{2}\sigma^{\prime} \left(\nabla p\right)^2\Big]\, \left(\nabla\cdot\delta \mathbf{u}\right) \nonumber\\
  &+& \int  d\mathbf{x}\,\delta \mathbf{u}(\mathbf{x},0)\cdot \nabla p(\mathbf{x},0)-\int d\mathbf{x} \,\delta \mathbf{u}(\mathbf{x},T)\cdot\nabla p(\mathbf{x},T) = 0.
   \label{P070}
\end{eqnarray}
We see that the bulk term in (\ref{P070}) and the second boundary term in time again yield Eq.~(\ref{d2}) and the boundary conditions (\ref{BCT}), respectively. Integrating the first boundary term by parts and assuming that $p(\mathbf{x},0)$ is sufficiently smooth, we see that the first boundary term vanishes because $\delta q(\mathbf{x},0)=\nabla\cdot\delta \mathbf{u} (\mathbf{x},0)=0$.

As one can notice, $q(\mathbf{x}\in\overline{\Omega},T)$ and $p(\mathbf{x}\in\Omega,T)$ are unknown \textit{a priori}. For a class of lattice gas models (which include the RWs, the SSEP and the KMP, among others), one can show that, as in the annealed setting, $p(\mathbf{x}\in\Omega,T)=-\infty$ in the particular case of void at $t=T$. This \textit{a priori} knowledge is convenient for solving the MFT equations, see Secs. \ref{RW} and \ref{num}.

One can also notice that, by fixing $q(\mathbf{x},0)=n$, we narrow the class of density variations in the problem of minimizing $\bar{S}$, in comparison with the annealed case. As a result, $\bar{S}$ in the quenched case is always greater than, or equal to, $\bar{S}$ in the annealed case. Therefore,  ${\cal P}_{\text{quenched}}\leq {\cal P}_{\text{annealed}}$.

\section{Microscopic theory for non-interacting random walkers}
\label{micro}

In this Appendix we outline a microscopic theory of void formation for the RWs, both in the quenched and annealed settings. This approach yields exact results for the void formation probability and for the expected final density profile. In the long time limit, exact results coincide with predictions of the MFT formalism. For simplicity, we present the exact results only in the one-dimensional case.

\subsection{Quenched setting}
\label{void_1d}

In the quenched setting, the initial condition is deterministic. We assume that, at $t=0$,  RWs occupy an infinite one-dimensional lattice, so that there are $n_m$ particles at site $m$.    We demand that the $[-L,L]$ interval of the lattice be empty at time $T$. Consider a single RW starting at $t=0$ at site $m$. This RW will be at site $j$ with probability $e^{-2T} I_{j-m}(2T)$,  where $I_k$ is the modified Bessel function (see, \textit{e.g.}, \cite{KRB10}). The probability for this RW to be outside the $[-L,L]$ interval at time $T$ is equal to the sum of arrival probabilities of this RW to all sites $j$ such that $|j|>L$:
\begin{equation}
\label{n=1}
e^{-2T}\sum_{|j|>L} I_{j-m}(2T)
\end{equation}
As all $n_m$ RWs on the same site $m$ are independent of each other, the probability that \emph{all} of them will be found outside the $[-L,L]$ interval at time $T$ is
\begin{equation}
\label{general_n}
\left[e^{-2T}\sum_{|j|>L} I_{j-m}(2T)\right]^{n_m}.
\end{equation}
Now, the RWs at other sites are also independent, so the probability $P(L, T)$ of the void formation at time $T$ is equal to an infinite product of probabilities (\ref{general_n}):
\begin{equation}
\label{hole_1d_inhom}
\mathcal{P}_{\text{quenched}}(L, T) = \prod_{-\infty<m<\infty}\left[e^{-2T}\sum_{|j|>L} I_{j-m}(2T)\right]^{n_m}.
\end{equation}
This is an exact result, valid for any deterministic initial configuration. In particular, for a deterministically homogeneous gas $n_m=n=\text{const}$, we obtain
\begin{equation}
\label{hole_1d}
\mathcal{P}_{\text{quenched}}(L, T) = \prod_{-\infty<m<\infty}\left[e^{-2T}\sum_{|j|>L} I_{j-m}(2T)\right]^{n}.
\end{equation}
The most interesting asymptotic behavior arises when $L\to\infty$ and $T\to\infty$, while the ratio $\ell = L/\sqrt{4T}$ remains finite. In this scaling region we can employ the asymptotic relation
\begin{equation}
\label{asymp_Bessel}
e^{-2T}I_m(2T)\simeq \frac{1}{\sqrt{4\pi T}}\,e^{-X^2}\,, \quad X=\frac{m}{\sqrt{4T}}\,,
\end{equation}
which holds when $T\to\infty$ and $m\to\infty$, while $X$ is finite.  This yields
\begin{equation}
e^{-2T}\sum_{|j|>L} I_{j-m}(2T) \simeq \frac{\text{erfc}(\ell-X)+\text{erfc}(\ell+X)}{2}\,.
\label{erfc1}
\end{equation}
Equation~(\ref{erfc1}) allows one to recast Eq.~\eqref{hole_1d} into the  scaling form (\ref{Squ}) and (\ref{P1_int1}).

We can also use the exact microscopic approach to determine the average gas density outside the void at  $t=T$. The probability that a RW starting at site $m$ is at site $x$ at time $T$ is $e^{-2T} I_{x-m}(2T)$. This probability is obtained by sampling over all evolution histories,  while we must consider a subset of histories when the RW ends up outside the $[-L,L]$ interval. The corresponding conditional  probability is $e^{-2T} I_{x-m}(2T)/\left[e^{-2T}\sum_{|j|>L} I_{j-m}(2T)\right]$, see (\ref{n=1}). Therefore, the average density outside the void at $t=T$ is given by
\begin{equation}
q(|x|>L,T) = n\, \sum_{m=-\infty}^\infty \frac{e^{-2T} I_{x-m}(2T)}{e^{-2T}\sum_{|j|>L} I_{j-m}(2T)}.
\end{equation}
In the limit of $x\to\infty$, $T\to\infty$  and finite $X = x/\sqrt{4T}$, we can use the asymptotic \eqref{asymp_Bessel}. As a result, the average gas density at $t=T$ acquires the scaling form (\ref{t=T}).

\subsection{Annealed setting}
\label{void_ann}

In the annealed setting the initial state of the gas exhibits equilibrium fluctuations. In this case, the evaluation of the probability of the void formation at time $T$ should be performed in two steps:
\begin{enumerate}
\item{Evaluate the probability of the void formation at time $T$ when starting from an arbitrary but fixed microscopic configuration (that is, for a quenched setting).}
\item{Average this probability over the microscopic equilibrium density distribution.}
\end{enumerate}
We have already performed step 1 of the calculation, and arrived at Eq.~(\ref{hole_1d_inhom}). It is convenient to temporarily deal with a finite lattice which includes $2N+1$ sites: $-N,-N+1, \dots, m, \dots, N-1,N$, where $N\gg L$.  The probability of void formation at time $T$ when starting from the microscopic configuration $(n_{-N}, n_{-N+1},\dots, n_m, \dots, n_{N-1}, n_{N})$ at time $t=0$ is
\begin{equation}
\label{hole_1d_ann_1}
\mathcal{P}_{\text{quenched}}(L, T) = \prod_{-N<m<N}\left[e^{-2T}\sum_{|j|>L} I_{j-m}(2T)\right]^{n_m}.
\end{equation}
We can now proceed to step 2. The (Poisson) microscopic equilibrium density distribution with a fixed total number of particles $K$ can be represented as
\begin{eqnarray}
&& \mathfrak{P}(n_{-N}, n_{-N+1},\dots, n_m, \dots, n_{N-1}, n_{N}) \nonumber \\
&& = \frac{K!\,\delta\left(\sum_{|m|<N}n_m-K\right)}{(2N)^K n_{-N}!\, n_{-N+1}! \,\dots \,n_m! \,\dots n_{N-1}! \,n_{N}!}\,,
\label{Poisson}
\end{eqnarray}
where $\delta$ is the Kronecker delta imposing the constraint of constant total number of particles:
$\sum_{|m|<N}n_m = K$. What is left is to average the quenched probability (\ref{hole_1d_ann_1}) over the distribution (\ref{Poisson}):
\begin{equation}
\mathcal{P}_{\text{annealed}} = \!\!\!\sum_{n_{-N}, \dots, n_N} \mathcal{P}_{\text{quenched}} (L,T)\, \mathfrak{P}(n_{-N}, \dots, n_m, \dots, n_{N}).
\label{exactann}
\end{equation}
The resulting expression is exact, but a more useful information can be extracted if we take the limits
of $T\to \infty$ and $L\to \infty$, keeping $\ell=L/\sqrt{4T}$ finite. Using Eq.~(\ref{erfc1}),
we obtain
\begin{equation}\label{hole_1d_ann_2}
\ln \mathcal{P}_{\text{annealed}} \simeq \sum_{|m|\leq N} n_m\,\ln \left[\frac{\text{erfc}(\ell-X)+\text{erfc}(\ell+X)}{2}\right],
\end{equation}
where $X=m/\sqrt{4T}$.  Using a continuous spatial coordinate $x$ instead of $m$ and a coarse-grained gas density $q(x,t=0)$ instead of $n_m$, we obtain
\begin{equation}
\label{hole_1d_ann_3}
\ln \mathcal{P}_{\text{annealed}} \simeq \int_{-\infty}^{\infty} dx \,q(x,0) \ln \chi(x,T),
\end{equation}
where $\chi(x,T)=\frac{1}{2}[\text{erfc}(\ell-X)+\text{erfc}(\ell+X)]$ with
$X=x/\sqrt{4T}$, and we have returned to an infinite system. Similarly, we can go over to the continuous limit in the exact microscopic density distribution (\ref{Poisson}). Employing Stirling's formula, $\ln (n_m!)\simeq n_m \, \ln n_m - n_m$, and replacing sums over $m$ by integrals over $x$, we obtain,  up
to a pre-exponent,
\begin{equation}\label{coarse}
\ln \mathfrak{P}[q(x)] \sim -\int_{-\infty}^{\infty}dx\,\left[q(x,0) \,\ln\frac{q(x,0)}{n}+n-q(x,0)\right].
\end{equation}
This is simply the Boltzmann-Gibbs cost of the macroscopic density profile $q(x,0)$ in the RWs model, see Eq.~(\ref{S0}) and Table 1.  Now Eq.~(\ref{exactann}) becomes a path integral:
\begin{equation}
\label{approxann}
   \mathcal{P}_{\text{annealed}} \sim \int \mathcal{D}q \,e^{-\int_{-\infty}^{\infty}dx\,\left(q \ln \chi-q \ln \frac{q}{n}-n+q\right)},
\end{equation}
which can be evaluated by the saddle-point method. The saddle point, in the functional space of $q(x,0)$, is found at $q(x,0)=n\, \chi(x,T)$ which coincides with Eq.~(\ref{q(x,0)}) for the most
likely initial density profile as predicted from the MFT formalism.
With this optimal initial density profile, Eq.~(\ref{approxann}) yields the equilibrium result
$\ln \mathcal{P}_{\text{annealed}} \simeq -2nL$.


\begin{thebibliography}{99}

\bibitem{LLStat}
    L. D. Landau and E. M. Lifshitz, {\it Statistical Physics} (New York: Pergamon Press, 1980).

\bibitem{Touchette} H. Touchette, Phys. Rep. \textbf{478}, 1 (2009).

\bibitem{D07}
     B. Derrida, J. Stat. Mech. P07023 (2007).

\bibitem{Jona}
    G. Jona-Lasinio, Prog. Theor. Phys. Suppl. {\textbf 184}, 262 (2010).

\bibitem{DG2009a}
    B. Derrida and A. Gerschenfeld, J. Stat. Phys. \textbf{136}, 1 (2009).

\bibitem{DG2009b}
    B. Derrida and A. Gerschenfeld, J. Stat. Phys. \textbf{137}, 978 (2009).

\bibitem{varadhan}
    S. Sethuraman and S.R.S. Varadhan, arXiv:1101.1479.

\bibitem{KM_var}
    P. L. Krapivsky and B. Meerson,  Phys. Rev. E \textbf{86}, 031106 (2012).

\bibitem{van}
   V. Lecomte, J. P. Garrahan, and F. van Wijland,  J. Phys. A: Math. Theor. \textbf{45}, 175001  (2012).

\bibitem{Abanov1} A. G. Abanov and V. E. Korepin, Nucl. Phys. B \textbf{647}, 565 (2002).

\bibitem{Abanov2}
    A. G. Abanov, ``Hydrodynamics of Correlated Systems. Emptiness Formation Probability
    and Random Matrices", in \textit{Application of Random Matrices in Physics}, edited by
    E. Brezin, V. Kazakov, D. Serban,  P. Wiegmann,  and A. Zabrodin,
    NATO Science Series II: Mathematics, Physics and Chemistry,
    Vol. 221 (Springer, Les Houches, 2004); arXiv:cond-mat/0504307.

\bibitem{Bertini}
    L. Bertini, A. De Sole, D. Gabrielli, G. Jona-Lasinio, and C. Landim,
    Phys. Rev. Lett. {\textbf 87}, 040601 (2001);  {\it ibid} {\textbf 94}, 030601 (2005);
    J. Stat. Phys. {\textbf 123}, 237 (2006); {\it ibid} {\textbf 135}, 857 (2009);
    J. Stat. Mech. (2007) P07014.

\bibitem{Tailleur}
    J. Tailleur, J. Kurchan, and V. Lecomte, Phys. Rev. Lett. \textbf{99}, 150602 (2007);
    J. Phys. A {\bf 41}, 505001 (2008).

\bibitem{FW84}
    M. I. Freidlin and A. D. Wentzell, {\it Random Perturbations of Dynamical Systems}
    (New York: Springer-Verlag, 1984).

\bibitem{EK}
    V. Elgart and A. Kamenev, Phys. Rev. E \textbf{70}, 041106 (2004).

\bibitem{MS2011}
    B. Meerson and P. V. Sasorov, Phys Rev. E \textbf{83}, 011129 (2011).

\bibitem{MSfronts}
    B. Meerson, P. V. Sasorov, and Y. Kaplan, Phys Rev. E \textbf{84}, 011147 (2011);
    B. Meerson and P. V. Sasorov, Phys Rev. E \textbf{84}, 030101(R) (2011).

\bibitem{Spohn}
     H. Spohn, {\it Large Scale Dynamics of Interacting Particles}
    (New York: Springer-Verlag, 1991).

\bibitem{L99}
    T. M. Liggett, {\it Stochastic Interacting Systems: Contact, Voter, and Exclusion Processes}
    (Springer, New York, 1999).

\bibitem{KL99}
     C. Kipnis and C. Landim, {\it Scaling Limits of Interacting Particle Systems}
     (Springer, New York,  1999).

\bibitem{SZ95}
     B. Schmittmann and R. K. P. Zia, \textit{Statistical Mechanics of Driven Diffusive Systems},
     in: {\it Phase Transitions and Critical Phenomena}, Vol.\ 17, eds.\ C. Domb and J. L. Lebowitz
     (Academic Press, London, 1995).

\bibitem{D98} B. Derrida, Phys.\ Rep.\ {\bf 301}, 65 (1998).

\bibitem{S00} G. M. Sch\"utz, \textit{Exactly Solvable Models for Many-Body Systems
  Far From Equilibrium}, in {\it Phase Transitions and Critical Phenomena},
  Vol.\ 19, eds.\ C. Domb and J. L. Lebowitz (Academic Press, London, 2000).

\bibitem{BE07} R. A. Blythe and M. R. Evans, J. Phys.\ A {\bf 40}, R333 (2007).

\bibitem{KRB10}
     P. L. Krapivsky, S. Redner, and E. Ben-Naim,
     {\it A Kinetic View of Statistical Physics} (Cambridge University Press, Cambridge,  2010).

\bibitem{KMP}
     C. Kipnis, C. Marchioro, and E. Presutti, J. Stat. Phys. {\textbf 27}, 65 (1982).

\bibitem{BGL}
    L. Bertini, D. Gabrielli,  and J. L. Lebowitz,  J. Stat. Phys. {\bf 121}, 843 (2005).

\bibitem{van_2}
    A. Imparato, V. Lecomte, and F. van Wijland,  Phys. Rev. E {\bf 80}, 011131 (2009).

\bibitem{Spitzer} F. Spitzer,  Adv. Math. \textbf{5}, 246 (1970).

\bibitem{Evans} M. R. Evans and T. Hanney, J. Phys. A: Math. Gen. \textbf{38}, R195 (2005).

\bibitem{monotone} This statement is valid for the ZRP only if the rate $\alpha(r) $ monotonically increases
with $r$.

\bibitem{ZRPproof}
    The derivation of relation $D(r)=\alpha^{\prime}(r)$ for the ZRP with $\alpha^{\prime}(r)>0$
    closely follows the one
    for RWs, see \cite{MS2011}.  One starts from the master equation for the multivariate
    probability distribution and applies WKB approximation employing $1/n_\mathbf{i}$ as a small
    parameter. An additional assumption is slow variation of the multivariate probability distribution
    (and therefore of the WKB action) along the lattice.

\bibitem{gradp}
   Since the momentum $p$ enters the MFT formalism only via $\nabla p$, it would be better
   to call the relaxation solutions
   those with $\nabla p =0$ and the activation solutions those with  $\nabla p \neq 0$.

\bibitem{Zeld}
     Ya. B. Zeldovich and Yu. P. Raizer,
     {\it Physics of Shock Waves and High-Temperature Hydrodynamic Phenomena}, Vols. I \& II
     (Academic Press, New York, 1967).

\bibitem{LL87}
     L. D. Landau and E. M. Lifshitz, {\it Fluid Mechanics}
    (Pergamon Press, New York, 1987).

\bibitem{B96}
     G.~I.~Barenblatt, {\it Scaling, Self-Similarity, and Intermediate
     Asymptotics} (Cambridge University Press, Cambridge, 1996).

\bibitem{pablo}
    P. I. Hurtado and P. L. Krapivsky, Phys. Rev. E \textbf{85}, 060103 (2012).

\bibitem{Stepanov}
    A. I. Chernykh and M. G. Stepanov, Phys. Rev. E \textbf{64}, 026306 (2001).

\bibitem{Korepin}
    V. E. Korepin, N. M. Bogoliubov, and A. G. Izergin, \textit{Quantum Inverse Scattering
    Method and Correlation Functions} (Cambridge University Press, Cambridge, UK, 1993).

\bibitem{vortices}
     In finite systems of spatial dimension $d\geq 2$,  the integral over
     $\nabla \cdot (p \boldsymbol{\omega})$ can give a non-trivial contribution to the action leading
     to new effects that are absent in vortex-free settings. One example is fluctuations of the
     stationary current through a finite-size slit in two-dimensional systems. These fluctuations
     can be dominated by vortices, see
     T. Bodineau, B. Derrida, and J. L. Lebowitz, J. Stat. Phys. \textbf{131}, 821 (2008).


\end{thebibliography}
\end{document}